\documentclass[11pt,onecolumn,draftclsnofoot]{IEEEtran}
\usepackage{amsmath,amssymb,graphicx}
\usepackage{algorithm,algorithmic}
\newfont{\bbb}{msbm10 scaled 500}

\newfont{\bb}{msbm10 scaled 1100}
\newcommand{\CC}{\mbox{\bb C}}
\newcommand{\RR}{\mbox{\bb R}}

\newcommand{\hv}{{\bf h}}

\newcommand{\mv}{{\bf m}}
\newcommand{\nv}{{\bf n}}

\newcommand{\uv}{{\bf u}}
\newcommand{\wv}{{\bf w}}
\newcommand{\vv}{{\bf v}}
\newcommand{\xv}{{\bf x}}
\newcommand{\yv}{{\bf y}}

\newcommand{\zerov}{{\bf 0}}

\newcommand{\Am}{{\bf A}}
\newcommand{\Bm}{{\bf B}}

\newcommand{\Dm}{{\bf D}}

\newcommand{\Hm}{{\bf H}}
\newcommand{\Id}{{\bf I}}

\newcommand{\Pm}{{\bf P}}
\newcommand{\Qm}{{\bf Q}}
\newcommand{\Rm}{{\bf R}}

\newcommand{\Tm}{{\bf T}}
\newcommand{\Um}{{\bf U}}
\newcommand{\Wm}{{\bf W}}
\newcommand{\Vm}{{\bf V}}
\newcommand{\Xm}{{\bf X}}

\newcommand{\Cc}{{\cal C}}

\newcommand{\Ic}{{\cal I}}

\newcommand{\Lc}{{\cal L}}

\newcommand{\Nc}{{\cal N}}
\newcommand{\Oc}{{\cal O}}

\newcommand{\Rc}{{\cal R}}
\newcommand{\Sc}{{\cal S}}

\newcommand{\Vc}{{\cal V}}

\newcommand{\Lambdam}{\hbox{\boldmath$\Lambda$}}

\newcommand{\diag}{{\hbox{diag}}}
\newcommand{\trace}{{\hbox{tr}}}

\renewcommand{\arg}{{\hbox{arg}}}

\renewcommand{\Im}{{\rm Im}}

\newcommand{\defines}{{\,\,\stackrel{\scriptscriptstyle \bigtriangleup}{=}\,\,}}

\newcommand{\htp}{^{\sf H}} 
\newcommand{\tp}{^{\sf T}}  
\def\LSB{\left[}        
\def\RSB{\right]}       
\def\LB{\left(}         
\def\RB{\right)}        

\newtheorem{definition}{Definition}
\newtheorem{lemma}{Lemma}
\newtheorem{cor}{Corollary}
\newtheorem{prop}{Proposition}

\newtheorem{remark}{\indent  \textit{Remark}}

\newtheorem{theorem}{Theorem}
\newtheorem{assumption}{\indent \bf A}

\begin{document}
\title{\Huge Random Beamforming over Correlated Fading Channels}
\author{
\IEEEauthorblockN{Jakob Hoydis, Romain Couillet, and M\'{e}rouane Debbah}
\thanks{J. Hoydis is with the Department of Telecommunications and the Alcatel-Lucent Chair on Flexible Radio, Sup\'{e}lec, 91192 Gif-sur-Yvette, France (e-mail: jakob.hoydis@supelec.fr).}
\thanks{R. Couillet is with the EDF Chair on System Sciences and the Energy Challenge, Centrale Paris-Sup\'ele, 91192 Gif-sur-Yvette, France (e-mail: romain.couillet@supelec.fr).}
\thanks{M. Debbah is with the Alcatel-Lucent Chair on Flexible Radio, Sup\'{e}lec, 91192 Gif-sur-Yvette, France (e-mail: merouane.debbah@supelec.fr).}}
\maketitle

\begin{abstract}
We study a multiple-input multiple-output (MIMO) multiple access channel (MAC) from several multi-antenna transmitters to a multi-antenna receiver. The fading channels between the transmitters and the receiver are modeled by random matrices, composed of independent column vectors with zero mean and different covariance matrices. Each transmitter is assumed to send multiple data streams with a random precoding matrix extracted from a Haar-distributed matrix. For this general channel model, we derive deterministic approximations of the normalized mutual information, the normalized sum-rate with minimum-mean-square-error (MMSE) detection and the signal-to-interference-plus-noise-ratio (SINR) of the MMSE decoder, which become arbitrarily tight as all system parameters grow infinitely large at the same speed. In addition, we derive the asymptotically optimal power allocation under individual or sum-power constraints. Our results allow us to tackle the problem of optimal stream control in interference channels which would be intractable in any finite setting. Numerical results corroborate our analysis and verify its accuracy for realistic system dimensions. Moreover, the techniques applied in this paper constitute a novel contribution to the field of large random matrix theory and could be used to study even more involved channel models. 
\end{abstract}

\clearpage
\section{Introduction}
The gains of having multiple antennas at the transmitter and receiver in wireless fading point-to-point channels are well established \cite{foschini1998,telatar1999}. It is known since the Telatar's seminal paper \cite{telatar1999} that when channel state information (CSI) is available at the transmitter, the optimal transmission strategy is to send independent data streams along the eigenmodes of the channel and to allocate power over these eigenmodes according to the water-filling principle \cite{coverbook}. If no CSI is available at the transmitter but the statistical properties of the channel are known, an optimal static power allocation which does not depend on the actual channel realizations and maximizes the ergodic mutual information can be found. Uniform power allocation is optimal when the channel entries are independent and identically distributed (i.i.d.\@) Gaussian \cite{telatar1999} or zero-mean symmetric \cite{palomarthesis}. It was shown in \cite{rhee2001} that the optimality of uniform power allocation also extends to the multiple-access channel. When not ergodic but outage capacity is considered, it was conjectured \cite{telatar1999} that allocating equal power to only a subset of the available transmit antennas is optimal. This conjecture was proved for the Gaussian multiple-input single-output (MISO) channel in \cite{jorswieck2007}.   

In the presence of co-channel interference, much less is known about the optimal transmission strategies. Recently, an exact expression of the ergodic mutual information of a Rayleigh fading multiple-input multiple-output (MIMO) system in the presence of different MIMO interferers with arbitrary transmit power levels was derived \cite{chiani2010}. Since more involved channel models are intractable by exact analysis, many works resort to asymptotic analyses where each transmitter and receiver is equipped with a large number of antennas. The authors of \cite{moustakas03} consider a doubly-correlated fading MIMO channel with correlated interference and derive asymptotically tight approximations of the mutual information and its variance. In \cite{allerton2010}, the asymptotic mutual information and its fluctuations are studied for  arbitrary fading channels with a variance profile and correlated noise. However, even in the asymptotic setting, the optimal transmit strategies in interference channels are in general unknown.

An important question in MIMO systems with co-channel interference is whether a transmitter should use all of its antennas to transmit independent data streams or whether it should restrict itself to a smaller number of streams or antennas in order to reduce the interference to other receivers. In general, this problem does not have a simple solution and the optimal number of antennas to be used (or streams to be sent) depends on the strength of the interference and, thus, on the cross-channel gains between the interferers. The authors of \cite{blum2003} pioneered this question, assuming that no CSI is available at the transmitters while full CSI is available at the receivers. Their main finding is that when the interference is weak, a transmitter should send independent streams with equal power from each of its antennas, but when the interference is strong, all power should be put into a single stream which is transmitted by a single antenna. In \cite{blum2002}, it was shown that optimizing the number of transmitted data streams is also helpful when CSI is available at the transmitter. Several later works \cite{vaze2009,louie2010,louie2011} have studied the same problem in the context of dense random ad hoc networks under different assumptions about the availability of CSI at the transmitters and receivers and the corresponding transmit and reception strategies. Surprisingly, the conclusions in all of these works are similar, confirming the optimality of single-stream transmissions in dense, interference-limited networks.

The aforementioned references share the underlying assumption that the channel matrices/vectors are composed of i.i.d.\@ elements without any form of correlation. Thus, the problem of how many antennas should be used for transmission and how many independent data streams should be sent are the same. With transmit antenna correlation, however, it makes a difference which antennas are selected for transmission and the question of the optimal number of antennas to be used becomes a combinatorial problem. To circumvent this issue, random isometric precoding can be used to mitigate transmit correlation. The remaining question is then how many orthogonal streams should be sent, using \emph{all} available antennas. This is the primal motivation of this paper, as our results allow to study the sum-rate of systems composed of multiple transmitter-receiver pairs, each applying random isotropic beamforming.
Random isotropic beamforming \cite{tse2002} is a well studied technique in multi-user MIMO communication systems and unitary precoders \cite{LOV05} are now proposed as limited feedback beamforming solutions in future wireless standards \cite{LEE09, HUA09}. Nevertheless, only few related works relying on tools from large random matrix theory have been published until today (e.g. \cite{couillet10b}) and this paper might stimulate further research in this area.

The contributions of this paper are as follows. We consider a MIMO multiple access channel (MAC) from several multi-antenna transmitters to a multi-antenna receiver. The transmitters are unaware of the channel realizations and send an arbitrary number of independent data streams using isometric random beamforming vectors. The receiver is assumed to be aware of all instantaneous channel realizations and beamforming vectors. We assume a very general channel model where the channel matrices are composed of independent, zero mean column vectors, each with a possibly different covariance matrix. This channel model allows to treat many classes of well-known channel models, such as matrices with a variance profile \cite{hachem07} as well as the Kronecker model \cite{couillet10}. Under these general assumptions, we derive deterministic approximations of the normalized mutual information, the normalized sum-rate with minimum-mean-square-error (MMSE) detection and the signal-to-interference-plus-noise-ratio (SINR) of the MMSE decoder, which become arbitrarily tight as all system parameters grow infinitely large at the same speed. The expressions are given as functions of a set of deterministic quantities which can be computed by a standard fixed-point algorithm which is proved to converge. Moreover, we derive the optimal power allocation under individual or sum-power constraints which can be computed by an iterative water-filling algorithm. We then apply these results to find the optimal number of independent streams to be transmitted in a $2\times2$ interference channel. Although the use of deterministic approximations in this context requires an exhaustive search over all possible stream-configurations, it is computationally much less expensive than Monte Carlo simulations. Extensions to more than two transmit-receive pairs and possible different objective functions, e.g. weighted sum-rate or sum-rate with MMSE decoding, are straightforward. Our numerical results show that the deterministic approximations are very tight for even small system dimensions. We further show that with random beamforming, it is optimal to (i) send as many independent data streams as transmit antennas and (ii) allocate power uniformly over the transmitted streams. For the interference channel, we find that at low SNR, it is optimal to use all streams while at high SNR, stream-control, i.e., transmitting less than the maximal number of streams, is beneficial. Apart from these practical applications, our work also constitutes a novel contribution to the field of random matrix theory as will be highlighted in Section~\ref{sec:literature}.

The remainder of this paper is organized as follows: In Section~\ref{sec:sys}, we present a detailed description of the system model and make several definitions of frequently used quantities. Section~\ref{sec:literature} summarizes recent results on large random matrix theory involving Haar distributed matrices, which will be extended in this paper. We present our main results in Section~\ref{sec:main} and show numerical results for practical applications such as optimal stream control for the MIMO interference channel in Section~\ref{sec:num}. The paper is concluded in Section~\ref{sec:con}. Related results, lemmas and the proofs of all theorems are provided in the appendix.

\vspace{10pt}\textbf{Notations}: Boldface lower and upper case symbols represent vectors and matrices, respectively. $\Id_N$ is the size-$N$ identity matrix and $\diag(x_1,\dots,x_N)$ is a diagonal matrix with elements $x_i$. The trace, transpose and Hermitian transpose operators are denoted by $\trace(\cdot)$, $(\cdot)\tp$ and $(\cdot)\htp$, respectively. 
The spectral norm of a matrix $\Am$ is denoted by $\lVert\Am\rVert$, and, for two matrices $\Am$ and $\Bm$, the notation $\Am\succ\Bm$ means that $\Am-\Bm$ is positive-definite. 
The notations $\Rightarrow$ and $\xrightarrow[]{\text{a.s.}}$ denote weak and almost sure convergence, respectively. We use $\Cc\Nc\left(\mv,\Rm\right)$ to denote the circular symmetric complex Gaussian distribution with mean $\mv$ and covariance matrix $\Rm$. We denote by $\RR_+$ the set $[0,\infty)$ and by $\CC_+$ the set $\{z\in\CC,\Im[z]>0\}$. 
Denote by $\Cc$ the set of continuous functions from $X\subset\CC$ to $Y\subset\CC$ and by $\Sc$ the class of functions $f$ analytic over $\CC\setminus\RR_+$, such that, for $z\in\CC_+$, $f\in\CC_+$, $zf\in\CC_+$ and $\lim_{y\to\infty}-{\bf i}yf({\bf i}y)<\infty$. Such functions are known to be \emph{Stieltjes transforms} of finite measures supported by $\RR_+$ (see e.g. \cite{couilletRMT}).

\section{System model}\label{sec:sys}
Consider the following discrete-time MIMO channel with output vector $\yv\in\CC^N$:
\begin{align}\label{eq:channel}
 \yv = \sum_{k=1}^K\Hm_k\Wm_k\Pm_k^{\frac12}\xv_k +\nv
\end{align}
where, for $k\in\{1,\dots,K\}$,
\begin{itemize}
\item[(i)] $\Hm_k\in \CC^{N\times N_k}$ is a random channel matrix whose $j$th column vector $\hv_{kj}\in\CC^{N}$ is modeled as
\begin{align}\label{eq:channelmodel}
\hv_{kj} = \Rm_{kj}^{\frac12}\uv_{kj},\qquad j=1,\dots,N_k 
\end{align}
where $\Rm_{kj}\in\CC^{N\times N}$ are Hermitian nonnegative definite matrices and the vectors $\uv_{kj}\in\CC^{N}$ have independent and identically distributed (i.i.d.) elements with zero mean, variance $1/N$ and finite moment of order $4+\epsilon$, for some common $\epsilon>0$,
\item[(ii)] $\Wm_k\in \CC^{N_k\times n_k}$ is a complex precoding matrix which contains $n_k\le N_k$ orthonormal columns of an $N_k\times N_k$ Haar-distributed random unitary matrix, 
\item[(iii)] $\Pm_k=\diag(p_{k1},\dots,p_{kn_k})\in\RR_+^{n_k\times n_k}$ is a nonnegative diagonal matrix,
\item[(iv)] $\xv_k\sim\Cc\Nc(0,\Id_{n_k})$ is the transmit vector of the $k$th transmitter,
\item[(v)] $\nv\sim\Cc\Nc(0,\rho\Id_{N})$ is a noise vector.
\end{itemize}

\vspace{10pt}\begin{remark}\label{rem:kronecker}
The statistical model \eqref{eq:channelmodel} generalizes several well-know channel models of interest (see \cite{wagner2011,hoydis2011} for examples). It comprises in particular the \emph{Kronecker} channel model with transmit and receive correlation matrices \cite{chuah02,couillet10}, where the matrices $\Hm_k$ are given as 
\begin{align}\label{eq:kronecker}
\Hm_k = \Rm_k^\frac12\Um_k\Tm_k^\frac12  
\end{align}
where $\Um_k\in\CC^{N\times N_k}$ is a random matrix whose elements are independent $\Cc\Nc(0,1/N)$ and $\Rm_k\in\CC^{N\times N}$ and $\Tm_k\in\CC^{N_k\times N_k}$ are covariance matrices. Since both $\Um_k$ and $\Wm_k$ are unitarily invariant, we can assume without loss of generality for the statistical properties of $\yv$ that $\Tm_k=\diag(t_{k1},\dots,t_{kN_k})$. Defining the matrices $\Rm_{kj}=t_{kj}\Rm_k$ for $j=1,\dots,N_k$, we fall back to the channel model in \eqref{eq:channelmodel}.
\end{remark}\vspace{10pt}

\vspace{10pt}\begin{remark}
Whenever the distribution of $\Hm_k$ is invariant by multiplications by unitary matrices from the right side (e.g. for \eqref{eq:kronecker} with $\Tm_k=\Id_{N_k}$), our channel model boils down to 
\begin{align*}
  \yv = \sum_{k=1}^K\Hm_k\Pm_k^{\frac12}\xv_k +\nv
\end{align*}
which has been studied in \cite{wagner2011} for the general case \eqref{eq:channelmodel} and in \cite{couillet10} for the Kronecker model \eqref{eq:kronecker}. The same holds for $n_k=N_k$ for all $k$ with uniform power allocation, i.e., $\Pm_k=\Id_{n_k}$, since $\Wm_k\Pm_k\Wm_k\htp=\Id_{N_k}$.
\end{remark}\vspace{10pt}

The next definitions will be of repeated use in the sequel. Let the matrix $\Bm_N\in\CC^{N\times N}$ be defined as 
\begin{align*}
\Bm_N = \sum_{k=1}^K \Hm_k\Wm_k\Pm_k\Wm_k\htp\Hm_k\htp.
\end{align*}
We denote by $I_N(\rho)$ the normalized mutual information of the channel \eqref{eq:channel}, given by \cite{coverbook}
\begin{align*}
 I_N(\rho) = \frac{1}{N}\log\det\LB \Id_N +\frac{1}{\rho}\Bm_N\RB
\end{align*}
expressed in $\text{nats}/\text{s}$. We further denote by $\gamma^N_{kj}$ the SINR at the output of the linear MMSE detector for the $j$th component of transmit vector $\xv_k$, which reads \cite{verdubook}
\begin{align*}
 \gamma^N_{kj} = p_{kj}\wv_{kj}\htp\Hm_k\htp\LB{\Bm_N}_{[kj]} + \rho\Id_N\RB^{-1}\Hm_k\wv_{kj}
\end{align*}
where ${\Bm_N}_{[kj]} = \Bm_N - p_{kj}\Hm_k\wv_{kj}\wv_{kj}\htp\Hm_k\htp$ and $\wv_{kj}$ is the $j$th column of $\Wm_k$.
We further define the normalized sum-rate with single-stream MMSE detection as
\begin{align*}
 R_N(\rho) = \frac1N\sum_{k=1}^K\sum_{j=1}^{n_k}\log\LB1+\gamma^N_{kj}\RB.
\end{align*}

The aim of this paper is to derive deterministic approximations of the quantities $I_N(\rho)$, $\gamma^N_{kj}$ and $R_N(\rho)$ which become almost surely arbitrarily tight as the dimensions of all involved matrices grow large. To make the definition of growth rigorous we need the following technical assumption:

For $k\in\{1,\dots,K\}$, let $\{N_k\}=\{N_k(N)\}$ and $\{n_k\}=\{n_k(N)\}$ be sequences of integers with ratios $c_k=c_k(N)=\frac{n_k}{N_k}$ and $\bar{c}_k=\bar{c}_k(N)=\frac{N_k}{N}$. The notation $N\to\infty$ should be understood from now on as  $N,N_1,\dots,N_k,n_1,\dots,n_K\to\infty$, such that  $0\le c_k \le 1$ and $0<\lim\inf_N \bar{c}_k \le \lim\sup_N \bar{c}_k<\infty$.
For all convergence results in this paper (as $N\to\infty$), the matrices $\Pm_k=\Pm_k(N)\in\RR_+^{n_k\times n_k}$, $\Rm_{kj}=\Rm_{kj}(N)\in\CC^{N\times N}$, $\Hm_k=\Hm_k(N)\in\CC^{N\times N_k}$ and $\Wm_k=\Wm_k(N)\in\CC^{N_K\times n_k}$ should be understood as families of (random) matrices with growing dimensions. Wherever this is clear from the context, we drop the dependence on $N$ to simplify the notations.

\section{State-of-the-art}\label{sec:literature}
Isotropically precoded systems with linear receivers have been studied in the asymptotic limit in several works. Relying on results from free probability theory, the authors of \cite{DEB03} investigate the asymptotic performance of the MMSE receiver for the channel model \eqref{eq:channel}, assuming $K=1$, $\Pm_1=\Id_{n_1}$ and $\Hm_1$ diagonal with i.i.d.\@ elements. Extensions of this work to frequency-selective fading channels with sub-optimal receivers were considered in \cite{HAC04}. Multi-carrier code-division multiple-access (MC-CDMA) with random i.i.d.\@ and isometric spreading sequences over Rayleigh fading channels, i.e., $K\ge 1$ and $\Hm_k$ diagonal with i.i.d.\@ complex Gaussian entries, was studied in \cite{PEA04} and approximate solutions of the SINR of the MMSE receiver were derived. In \cite{PEA06}, an expression of the asymptotic spectral efficiency for the same model was presented. In a later work \cite{COU09d}, DS-CDMA over flat-fading channels was considered, i.e., $K\ge 1$, $n_k=N$ and $\Hm_k=\Id_N$ for all $k$. The authors derive in particular deterministic approximations of the Shannon- and $\eta$-transform, exploiting the asymptotic freeness \cite[Section 3.5]{couilletRMT} of the matrices $\Wm_k\Pm_k\Wm_k\htp$. They further present a sum-rate maximizing power-allocation algorithm. The exact asymptotic SINR of the MMSE receiver for the general channel model \eqref{eq:channel} with deterministic, jointly diagonalizable matrices $\Hm_k$\footnote{The matrices $\Hm_1,\dots,\Hm_K$ are jointly diagonalizable if there exists a unitary matrix $\Vm$ such that $\Vm\Hm_k\Hm_k\htp\Vm\htp$ is diagonal for all $k$.} was found in \cite{PEA08} via incremental matrix expansions.

In \cite{couillet10b}, the channel model \eqref{eq:channel} with arbitrary deterministic matrices $\Hm_k$ was considered and deterministic approximations of the Stieltjes transform, the normalized mutual information and the SINR of the MMSE receiver, which are asymptotically almost surely tight, were established. As the results of the current paper build heavily on their work, we will restate the main theorems from \cite{couillet10b}. The aim of this paper is to extend these results to the case where the matrices $\Hm_k$ are random and modeled according to \eqref{eq:channelmodel}.

The first theorem in \cite{couillet10b} introduces a set of $2K$ implicit equations whose unique solution defines some quantities $\bar{e}_1,\dots,\bar{e}_K,e_1,\dots,e_K$. It will turn out that the normalized mutual information as well as the SINR of the MMSE receiver can be expressed as functions of these quantities. Also a fixed-point algorithm for the computation of  $\bar{e}_1,\dots,\bar{e}_K,e_1,\dots,e_K$ is provided in \cite{couillet10b}, which is guaranteed to converge to the correct solution.

\vspace{10pt}\begin{theorem}[{\cite[Theorem 1]{couillet10b}}] \label{th:fundequdet} 
For $k\in\{1,\dots,K\}$, let $\Pm_k\in \RR_+^{n_k\times n_k}$ be a diagonal matrix and let $\Hm_k\in\CC^{N\times N_k}$. Define $c_k= \frac{n_k}{N_k}$ and $\bar{c}_k= \frac{N_k}{N}$. Then, the following system of implicit equations in $(\bar{e}_k(z),\ldots,\bar{e}_K(z))$:
\begin{align*} 
  \bar{e}_k(z) &= \frac1N\trace \Pm_k\LB e_k(z) \Pm_k + [\bar{c}_k-e_k(z) \bar{e}_k(z)] \Id_{n_k} \RB^{-1} \\
  e_k(z) &= \frac1N\trace \Hm_k\Hm_k\htp\LB\sum_{j=1}^K \bar{e}_j(z) \Hm_j\Hm_j\htp -z\Id_N \RB^{-1}
\end{align*}
has a unique solution $(\bar{e}_1(z),\dots,\bar{e}_K(z))\in\Cc^K$ satisfying $(e_1(z),\dots,e_K(z))\in\Sc^K$ and, for $z<0$ and for all $k$, $0\le \bar{e}_k(z) < c_k\bar{c}_k/e_k(z)$.
\end{theorem}\vspace{10pt}

The next theorem utilizes the quantities provided by Theorem~\ref{th:fundequdet} to establish a deterministic equivalent to the normalized mutual information assuming that the matrices $\Hm_k$ are random. However, no particular random matrix model is specified and the only condition is the almost surely bounded spectral radius of the matrices  $\Hm_k\Hm_k\htp$.

\vspace{10pt}\begin{theorem}[{\cite[Theorem 3]{couillet10b}}]\label{th:logdetdet}
For $k\in\{1,\dots,K\}$, let $\Pm_k\in \RR_+^{n_k\times n_k}$ be a diagonal matrix with spectral norm bounded uniformly along $n_k$ and $\Wm_k\in\CC^{N_k\times n_k}$ be $n_k\leq N_k$ columns of a unitary Haar distributed random matrix. Consider $\Hm_k\in\CC^{N\times N_k}$ a random matrix such that $\Rm_k\defines \Hm_k\Hm_k\htp\in\CC^{N\times N}$ has uniformly bounded spectral norm along $N$, almost surely. Define $c_k= \frac{n_k}{N_k}$ and $\bar{c}_k= \frac{N_k}{N}$ and denote $\Bm_N = \sum_{k=1}^K \Hm_k\Wm_k\Pm_k\Wm_k\htp\Hm_k\htp$. Further, let $\rho>0$ and define
$\Ic_{N}(\rho)=\frac1N\log\det(\Id_N+\frac{1}{\rho}\Bm_N)$. Then, 
\begin{align*} 
     \Ic_{N}(\rho) - \bar{\Ic}_N(\rho) \xrightarrow[N\to\infty]{\text{a.s.}} 0
\end{align*}
where 
\begin{align*} 
\bar{\Ic}_N(\rho) &= \frac1N\log\det\left(\Id_N + \frac{1}{\rho}\sum_{k=1}^K \bar{e}_k\Rm_k\right)\\ &\qquad + \sum_{k=1}^K \left[\frac1N\log\det\left([\bar{c}_k-e_k\bar{e}_k]\Id_{n_k} + e_k\Pm_k \right) + (1-c_k)\bar{c}_k \log(\bar{c}_k-e_k \bar{e}_k) - \bar{c}_k \log(\bar{c}_k) \right]
\end{align*}
and where $e_k=e_k(-\rho)$, $\bar{e}_k=\bar{e}_k(-\rho)$ for all $k$ are given by Theorem~\ref{th:fundequdet}.
\end{theorem}\vspace{10pt}

In \cite[Theorem 4]{couillet10b}, also a deterministic approximation of the SINR at the output of the MMSE receiver is provided. Again, this theorem builds upon the solutions to the fundamental equations in Theorem~\ref{th:fundequdet} and will not be stated here for brevity.
In the next section, we will derive analogous results to the above theorems under the assumption that $\Hm_k$ are random and modeled by \eqref{eq:channelmodel}.

\clearpage
\section{Main Results}\label{sec:main}
Our first result is a generalization of Theorem~\ref{th:fundequdet} to the channel model \eqref{eq:channelmodel}. The quantites $g_k(\rho)$, $\bar{g}_k(\rho)$ and $\delta_{kj}(\rho)$, which are defined in the next theorem as the unique solution to a set of implicit equations, can be seen as the counterparts of $e_k$ and $\bar{e}_k$ in Theorem~\ref{th:fundequdet}. They will be similarly used to provide deterministic approximations of the mutual information and of the SINR of the MMSE detector. All proofs can be found in Appendix~\ref{app:proofs}.

\vspace{10pt}\begin{theorem}[Fundamental equations]\label{th:fundeq}
For $k\in\{1,\dots,K\}$, let $\Pm_k\in \RR_+^{n_k\times n_k}$ be a diagonal matrix and, for $j\in \{1,\dots,N_k\}$, let $\Rm_{kj}\in\CC^{N\times N}$ be a nonnegative-definite Hermitian matrix. Let $\rho>0$ and define $c_k=\frac{n_k}{N_k}$ and $\bar{c}_k=\frac{N_k}{N}$. Then, the following system of implicit equations:
 \begin{align*}
  \bar{g}_k(\rho) & = \frac1N\trace\Pm_k\Big( g_k(\rho)\Pm_k + \LSB\bar{c}_k-g_k(\rho)\bar{g}_k(\rho)\RSB\Id_{n_k}\Big)^{-1}\\
g_k(\rho) & = \frac1N\sum_{j=1}^{N_k}\frac{\delta_{kj}(\rho)}{1+\bar{g}_k(\rho)\delta_{kj}(\rho)}\\
\delta_{kj}(\rho) &= \frac1N\trace\Rm_{kj}\LB\frac{1}{N}\sum_{k=1}^K\sum_{j=1}^{N_k}\frac{\bar{g}_k(\rho)\Rm_{k,j}}{1+\bar{g}_k(\rho)\delta_{kj}(\rho)}+\rho\Id_N\RB^{-1}
\end{align*}
has a unique solution satisfying $\delta_{kj}(\rho)\ge0$, $g_k(\rho)\ge 0$  and $0\le \bar{g}_k(\rho)< c_k\bar{c}_k/g_k(\rho)$ for all $k,j$. 
Moreover, this solution is given explicitly by the following fixed-point algorithm:
\begin{align*}
\quad g_k(\rho) &= \lim_{t\to\infty} g^{(t)}_k(\rho),\quad\bar{g}_k(\rho) = \lim_{t\to\infty} \bar{g}^{(t)}_k(\rho),\quad \delta_{kj}(\rho) = \lim_{t\to\infty} \delta_{kj}^{(t)}(\rho)
\end{align*}
where 
\begin{align*}
\bar{g}^{(t)}_k(\rho) &= \lim_{l\to\infty}\bar{g}^{(t,l)}_k(\rho), \quad \delta_{kj}^{(t)}(\rho) = \lim_{l\to\infty}\delta_{kj}^{(t,l)}(\rho)\\
g_k^{(t)}(\rho)&=\frac1N\sum_{j=1}^{N_k}\frac{\delta_{kj}^{(t)}(\rho)}{1+\bar{g}_k^{(t-1)}(\rho)\delta_{kj}^{(t)}(\rho)}\\
\bar{g}^{(t,l)}_k(\rho)&= \frac1N\trace\Pm_k\LB g_k^{(t-1)}(\rho)\Pm_k+\LSB\bar{c}_k-g_k^{(t-1)}(\rho)\bar{g}_k^{(t,l-1)}(\rho)\RSB\Id_{n_k}\RB^{-1}\\
\delta_{kj}^{(t,l)}(\rho) &= \frac1N\trace\Rm_{kj}\LB\frac{1}{N}\sum_{k=1}^K\sum_{j=1}^{N_k}\frac{\bar{g}_k^{(t-1)}(\rho)\Rm_{k,j}}{1+\bar{g}_k^{(t-1)}(\rho)\delta_{kj}^{(t,l-1)}(\rho)}+\rho\Id_N\RB^{-1}
\end{align*}
with the initial values $\delta_{kj}^{(t,0)}(\rho)=1/\rho$, $\bar{g}_k^{(t,0)}\in[0,c_k\bar{c}_k/g_k^{(t-1)}(\rho))$, $\bar{g}_k^{(0)}=0$ and $g_k^{(0)}=0$ for all $k,j$.
\end{theorem}\vspace{10pt}

Before we present our results on the normalized mutual information and on the SINR of the MMSE receiver, we need the following two assumptions about the covariance matrices $\Rm_{kj}$ and the power allocation matrices $\Pm_k$:

\vspace{10pt}\begin{assumption}\label{as:finset}
$\Rm_{kj}\in\Rc_N$ for all $k,j$, where  $\Rc_N=\{\tilde{\Rm}_m, m=1,\dots,M\}$ is a family of $M$ Hermitian nonnegative-definite $N\times N$ matrices, satisfying 
$ \max_m\{\lim\sup_N\lVert\tilde{\Rm}_m\rVert\}\le R < \infty.$
\end{assumption}

\vspace{10pt}\begin{assumption}\label{as:limpow}
 For all $k$, $\lim\sup_N\lVert\Pm_k\rVert\le P < \infty.$ 
\end{assumption}

\vspace{10pt}\begin{remark}
While Assumption \textbf{A}~\ref{as:finset} is sufficient to ensure that the matrices $\Hm_k\Hm_k\htp$ have almost surely bounded spectral norm \cite[Proof of Theorem 3]{hoydis2011}, Assumption \textbf{A}~\ref{as:limpow} is necessary to ensure that no transmitter allocates an increasing amount of power to any of the streams as $N\to\infty$.
\end{remark}\vspace{10pt}

The next theorem extends Theorem~\ref{th:logdetdet} to random matrices $\Hm_k$ (as given by \eqref{eq:channelmodel}) and provides additionally a deterministic approximation of the normalized \emph{ergodic} mutual information.

\vspace{10pt}\begin{theorem}[Mutual information]\label{th:mutinf}
Assume that Assumptions \textbf{A}~\ref{as:finset} and \textbf{A}~\ref{as:limpow} hold true. Let $\rho>0$ and let $\bar{g}_k=\bar{g}_k(\rho)$,   $g_k=g_k(\rho)$ and $\delta_{kj}=\delta_{kj}(\rho)$ for all $k,j$ be defined as in Theorem~\ref{th:fundeq}. 
Then, 
\begin{align*}
 (i)\quad I_N(\rho) - \bar{I}_N(\rho) \xrightarrow[N\to\infty]{\text{a.s.}} 0,\qquad (ii)\quad \mathbb{E} I_N(\rho) - \bar{I}_N(\rho) \xrightarrow[N\to\infty]{} 0  
\end{align*}
where 
 \begin{align}\nonumber
 \bar{I}_N(\rho) &= \bar{V}_N(\rho) + \frac1N\sum_{k=1}^K\log\det\LB\LSB\bar{c}_k-g_k\bar{g}_k\RSB\Id_{n_k}+g_k\Pm_k\RB + \sum_{k=1}^K(1-c_k)\bar{c}_k\log(\bar{c}_k-g_k\bar{g}_k)-\bar{c}_k\log(\bar{c}_k)\\\label{eq:mutinfeq}
\bar{V}_N(\rho) &= \frac1N\log\det\LB\Id_N +\frac1\rho\frac{1}{N}\sum_{k=1}^K\sum_{j=1}^{N_k}\frac{\bar{g}_k(z)\Rm_{k,j}}{1+\bar{g}_k(z)\delta_{kj}(\rho)} \RB-\sum_{k=1}^K\bar{g}_kg_k + \frac1N\sum_{k=1}^K\sum_{j=1}^{N_k}\log\LB1+\bar{g}_k\delta_{kj}\RB .
\end{align}
\end{theorem}

\vspace{10pt}\begin{remark}
One can also consider an equivalent model where the matrices $\Pm_k$ are extended to $N_k\times N_k$ matrices by adding $N_k-n_k$ zeros to their main diagonal. This leads to $c_k=1$ for all $k$ and the expression of $\bar{I}_N(\rho)$ in \eqref{eq:mutinfeq} can be simplified accordingly.
\end{remark}\vspace{10pt}

Similarly, we obtain an extension to \cite[Theorem 4]{couillet10b} for the asymptotic SINR at the output of the MMSE receiver based on the fundamental equations in Theorem~\ref{th:fundeq}.

\vspace{10pt}\begin{theorem}[SINR of the MMSE detector]\label{th:sinr}
 Assume that Assumptions \textbf{A}~\ref{as:finset} and \textbf{A}~\ref{as:limpow} hold true. Let $\rho>0$ and define $g_k=g_k(\rho)$ and $\bar{g}_k=\bar{g}_k(\rho)$, as  given by Theorem~\ref{th:fundeq}. Then, 
\begin{align*}
 \gamma^N_{kj} - \bar{\gamma}^N_{kj} \xrightarrow[N\to\infty]{\text{a.s.}} 0
\end{align*}
where 
\begin{align*}
 \bar{\gamma}^N_{kj}=\frac{p_{kj} g_k}{\bar{c}_k-g_k\bar{g}_k}.
\end{align*}
\end{theorem}
\vspace{10pt}

The next corollary from Theorem~\ref{th:sinr} provides an asymptotically tight approximation of the normalized (ergodic) sum-rate with single-stream MMSE detection:

\vspace{10pt}\begin{cor}\label{cor:MMSE}
Assume that Assumptions \textbf{A}~\ref{as:finset} and \textbf{A}~\ref{as:limpow} hold true and let $\bar{\gamma}^N_{kj}$ be as defined in Theorem~\ref{th:sinr}. Then,
\begin{align*}
 (i)\quad R_N(\rho) - \bar{R}_N(\rho) \xrightarrow[N\to\infty]{\text{a.s}} 0,\qquad  (ii)\quad \mathbb{E}R_N(\rho) - \bar{R}_N(\rho) \xrightarrow[N\to\infty]{} 0
\end{align*}
where 
\begin{align*}
 \bar{R}_N(\rho) = \frac1N\sum_{k=1}^K\sum_{j=1}^{n_k}\log\LB1+\bar{\gamma}^N_{kj}\RB.
\end{align*}
\end{cor}\vspace{10pt}

Our last result is the asymptotically optimal power allocation which maximizes the normalized ergodic mutual information under individual and sum-power constraints:

\vspace{10pt}\begin{prop}[Optimal power allocation]\label{prop:pow}
 Let $\rho>0$ and $\bar{I}(\rho)$ be defined as in Theorem~\ref{th:mutinf} and let $P,P_1,\dots,P_K\ge 0$. Then, the solution to the following optimization problem:
\begin{align*}
 (\bar{\Pm}_1^*,\dots,\bar{\Pm}_K^*)\ =&\ \arg\max_{\Pm_1,\dots,\Pm_K} \bar{I}_N(x)\\
&\quad\text{s.t. } \begin{cases} \frac1{n_k}\trace\Pm_k \le P_k\ \forall k &(I)\\ \sum_{k=1}^K\frac1{n_k}\trace\Pm_k \le P & (II) \end{cases} 
\end{align*}
is given by
\begin{align*}
 \bar{p}_{kj}^* = \begin{cases} P_k & (I)\\
                   \LB\bar{g}_k^* - \frac{\bar{c}_k}{g_k^*} + \frac{c_k\bar{c}_k}{\lambda}\RB^{+} & (II)
                  \end{cases}
\end{align*}
for all $k,j$, where $\bar{\Pm}_k^*=\diag(\bar{p}_{kj}^*, j=1,\dots,n_K)$,  $g_k^*,\bar{g}_k^*$ are given by Theorem~\ref{th:fundeq} for $\Pm_k=\bar{\Pm}_k^*$,  and $\lambda$ in $(II)$ is chosen such that $\sum_{k=1}^K\frac1{n_k}\trace\bar{\Pm}_k^* = P$.
Moreover, let
\begin{align}\nonumber
 ({\Pm}_1^*,\dots,{\Pm}_K^*)\ =&\ \arg\max_{\Pm_1,\dots,\Pm_K} \mathbb{E} {I}_N(x)\\\label{eq:optIn}
&\quad\text{s.t. } \begin{cases} \frac1{n_k}\trace\Pm_k \le P_k\ \forall k &(I)\\ \sum_{k=1}^K\frac1{n_k}\trace\Pm_k \le P & (II) \end{cases}  
\end{align}
and assume that Assumptions \textbf{A}~\ref{as:finset} and \textbf{A}~\ref{as:limpow} hold true, then,
\begin{align*}
 \mathbb{E}I_N({\Pm}_1^*,\dots,{\Pm}_K^*) - \bar{I}_N(\bar{\Pm}_1^*,\dots,\bar{\Pm}_K^*) \xrightarrow[N\to\infty]{\text{a.s.}}0.
\end{align*}
\end{prop}

\vspace{10pt}\begin{remark}
The optimal power allocation matrices $\bar{\Pm}_k^*$ under a sum-power constraint $(II)$ can be computed by the iterative water-filling algorithm below. Although we cannot prove the sure convergence of this algorithm (see \cite[Remark 2]{couillet10} and \cite{dumont2010} for a related discussion), we know that if it converges, it achieves the correct solution. In our simulations, we could not create a case in which it did not converge.
\begin{algorithm}
\caption{Iterative water-filling algorithm}
\label{alg:wf}
 \begin{algorithmic}[1]
 \STATE Let $\epsilon>0$, $t=0$ and $\bar{p}_{kj}^{(0)}=P_k$ for all $k,j$.
 \REPEAT
 \STATE For all $k$, compute $\bar{g}_k^{(t)}$ and $g_k^{(t)}$ according to Theorem~\ref{th:fundeq} for the matrices $\Pm_k=\diag\LB \bar{p}_{kj}^{(t)}\RB$.
 \STATE For all $k,j$, calculate  $\bar{p}_{kj}^{(t+1)}=\LB\bar{g}_k^{(t)} - \frac{\bar{c}_k}{g_k^{(t)}} + \frac{c_k\bar{c}_k}{\lambda}\RB^{+}$, with $\lambda$ such that $\sum_{k=1}^K\frac1{n_k}\sum_{j=1}^{n_k}\bar{p}_{kj}^{(t+1)}=P$.
 \STATE $t=t+1$
 \UNTIL{$\max_{k,j} |\bar{p}_{k,j}^{(t)}-\bar{p}_{k,j}^{(t-1)}|\le\epsilon$} 
\end{algorithmic}
\end{algorithm}
\end{remark}

\vspace{10pt}\begin{remark}
The optimal power allocation also shows that sending as many independent data streams as transmit antennas is optimal to maximize the ergodic mutual information. This might not be the case in interference channels as will be discussed later on. 
\end{remark}

\vspace{10pt}\begin{remark}
 For the special case $K$=1, $\Pm_1=\Id_{n_1}$, $N_1=n_1=N$ and  $\Rm_{1j}=\Id_N$ for all $j$, the set of implicit equations in Theorem~\ref{th:fundeq} reduces to:
\begin{align*}
\bar{g}(\rho) &= \frac{1}{1-g(\rho)\bar{g}(\rho) + g(\rho)},\quad
g(\rho) = \frac{\delta(\rho)}{1+\bar{g}(\rho)\delta(\rho)},\quad
\delta(\rho)= \frac{1}{\frac{\bar{g}(\rho)}{1+\bar{g}(\rho)\delta(\rho)}+\rho}.
\end{align*}
Note that 
\begin{align*}
 0 &= 1 - \frac{1-g(\rho)\bar{g}(\rho) + g(\rho)}{1-g(\rho)\bar{g}(\rho) + g(\rho)}= 1 - \LSB1-g(\rho)\bar{g}(\rho)\RSB\bar{g}(\rho) -g(\rho)\bar{g}(\rho) =\LSB1-g(\rho)\bar{g}(\rho)\RSB(1-\bar{g}(\rho))
\end{align*}
which implies $\bar{g}(\rho)=1$ since $1-g(\rho)\bar{g}(\rho)>0$ by definition. Thus, the last equations further simplify to
\begin{align*}
g(\rho) = \frac{\delta(\rho)}{1+\delta(\rho)},\qquad \delta(\rho)= \frac{1}{\frac{1}{1+\delta(\rho)}+\rho}
\end{align*}
which has a unique solution satisfying $\delta(\rho)\ge0$ and that can be given in closed-form:
\begin{align*}
 \delta(\rho) = \frac{-1+\sqrt{1+\frac{4}{\rho}}}{2}.
\end{align*}
Notice that $\delta(\rho)$ is the Stieltjes transform of the Mar\u{c}enko-Pastur law \cite[Eq. (2.20)]{couilletRMT} evaluated on the negative real axis. This result is consistent with our expectations since $\Bm=\Um\Um\htp$, where $\Um\in\CC^{N\times N}$ has i.i.d.\@ entries with zero mean and variance $1/N$. Moreover, the expression of the normalized asymptotic mutual information as given in Theorem~\ref{th:mutinf} reduces to
\begin{align*}
 \bar{I}_N(\rho) = \bar{V}_N(\rho) = \log\LB1+\delta(\rho)+1/\rho\RB - \frac{\delta(\rho)}{1+\delta(\rho)}
\end{align*}
which is consistent with the asymptotic spectral efficiency of a Rayleigh-fading $N\times N$ MIMO channel \cite[Eq. (9)]{verdu99} (see also \cite[Section 12.2.2]{couilletRMT}). Equivalently, the asymptotic SINR of the MMSE detector and the associated normalized sum-rate can be given as (cf. \cite[Proposition VI.1]{verdu99}):
\begin{align*}
 \bar{\gamma}_j^N = \delta(\rho),\qquad \bar{R}_N(\rho)=\log(1+\delta(\rho)).
\end{align*}
\end{remark}
\vspace{10pt}

In the following section, we will present some applications of the previously derived results in the context of multiple-access and interference channels.

\section{Numerical Examples}\label{sec:num}
\subsection{Multiple Access Channel}\label{sec:MAC}
\begin{figure}
\centering \includegraphics[width=0.45\textwidth]{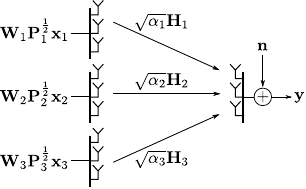}
\caption{MIMO MAC from three transmitters ($k=1,2,3$) with $N_k$ antennas to a receiver with $N$ antennas. Each transmitter sends $n_k$ streams with precoding matrix $\Wm_k$ and power allocation $\Pm_k$ over the channel $\sqrt{\alpha_k}\Hm_k$.\label{fig:MAC}}
\end{figure}

We consider a MAC from three transmitters to a single receiver as shown in Fig.~\ref{fig:MAC}. The channel from each transmitter is modeled by the Kronecker model (see Remark~\ref{rem:kronecker}) with individual transmit and receive covariance matrices $\Tm_k$ and $\Rm_k$ and we assume additionally a different path loss $\alpha_k>0$ on each link. The received signal vector $\yv$ for this model reads
\begin{align*}
 \yv = \sum_{k=1}^3 \sqrt{\alpha_k} \Rm_k^{\frac12}\Um_k\Tm_k^{\frac12}\Wm_k\Pm_k^{\frac12}\xv_k + \nv.
\end{align*}
We create the correlation matrices according to a generalization of Jake's model, as recently introduced in \cite{couillet10}, where the elements of $\Tm_k$ and $\Rm_k$ are given as
\begin{align}\nonumber
 \LSB\Tm_k\RSB_{ij} &= \frac{1}{\theta^{t,k}_\text{max}-\theta^{t,k}_\text{min}}\int_{\theta^{t,k}_\text{min}}^{\theta^{t,k}_\text{max}}\exp\LB\frac{{\bf{i}} 2\pi}{\lambda}d^{t,k}_{ij}\cos\LB\theta\RB\RB d\theta\\\label{eq:macchnmodel}
\LSB\Rm_k\RSB_{ij} &= \frac{1}{\theta^{r,k}_\text{max}-\theta^{r,k}_\text{min}}\int_{\theta^{r,k}_\text{min}}^{\theta^{r,k}_\text{max}}\exp\LB\frac{{\bf{i}} 2\pi}{\lambda}d^r_{ij}\cos\LB\theta\RB\RB d\theta
\end{align}
where $(\theta^{t,k}_\text{min},\theta^{t,k}_\text{max})$ and $(\theta^{r,k}_\text{min},\theta^{r,k}_\text{max})$ determine the solid angles over which useful signal power for the $k$th transmitter is radiated or received, $d^{t,k}_{ij}$ and $d^{r}_{ij}$ are the distances between the antenna elements $i$ and $j$ at the $k$th transmitter and receiver, respectively, and $\lambda$ is the signal wavelength. We assume uniform power allocation, i.e., $\Pm_k=\Id_{n_k}$ for all $k$, and define the signal-to-noise ratio $\text{SNR}=1/\rho$. The explicit choices of all other parameters are summarized in Table~\ref{tab:param}. 

\begin{figure}
\centering
 \includegraphics[width=0.55\textwidth]{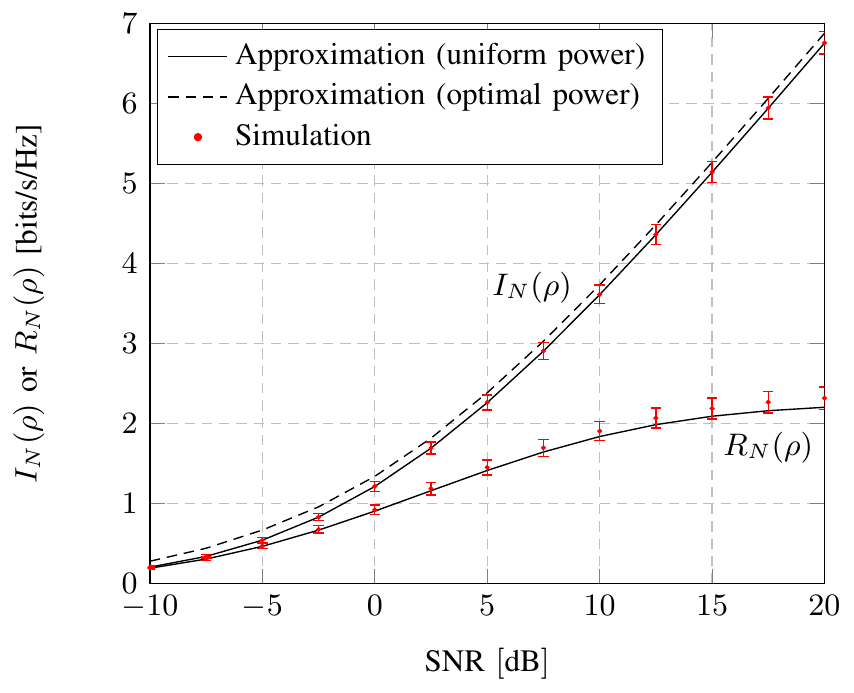}
\caption{Comparison of the average normalized mutual information $I_N(\rho)$ and the normalized rate with MMSE decoding $R_N(\rho)$ with their deterministic approximations $\bar{I}_N(\rho)$ and $\bar{R}_N(\rho)$. Error bars represent one standard deviation in each direction.\label{fig:MAC_mutinf}}
\end{figure}

Fig.~\ref{fig:MAC_mutinf} compares the normalized mutual information $I_N(\rho)$ and the normalized rate with MMSE decoding $R_N(\rho)$, averaged over $10,000$ different realizations of the matrices $\Hm_k$ and $\Wm_k$, against their deterministic approximations $\bar{I}_N(\rho)$ and $\bar{R}_N(\rho)$. Although we have chosen small dimensions for all matrices (see Table~\ref{tab:param}), the match between both results is almost perfect. Also the fluctuations of  $I_N(\rho)$ and $R_N(\rho)$ are rather small as can be seen from the error bars representing one standard deviation in each direction. The figure further illustrates the gains of optimal power allocation with a sum-power constraint $(II)$, where we have chosen $P=\sum_{k=1}^3\frac1{n_k}\trace\Id_{n_k}=3$ to achieve the same total transmitted power as for uniform power allocation, i.e., $\Pm_k=\Id_{n_k}$.

\begin{table}
\renewcommand{\arraystretch}{1.3}
\caption{Simulation parameters for Fig.~\ref{fig:MAC_mutinf}: $N=10$, $d^{r}_{ij}=8\lambda(i-j)$}
\label{tab:param}
\centering
\begin{tabular}{c|cccccccc}
$k$ & $N_k$ & $n_k$ & $\theta^{t,k}_\text{min}$ & $\theta^{t,k}_\text{max}$ & $\theta^{r,k}_\text{min}$ & $\theta^{r,k}_\text{max}$ & $d^{t,k}_{ij}$& $\alpha_k$ \\
\hline\hline
1 & 10 & 8 & $0$      & $\pi/2$ & $-\pi/4$ & $0$     & $4\lambda(i-j)$ & $1$   \\\hline
2 & 5  & 4 & $-\pi/4$ & $\pi/4$ & $0$      & $\pi/3$ & $4\lambda(i-j)$ & $1/2$ \\\hline
3 & 5  & 4 & $-\pi/2$ & $0$     & $-\pi/3$ & $\pi/3$ & $4\lambda(i-j)$ & $1/2$ \\\hline
\end{tabular}
\end{table}

\subsection{Stream-control in interference channels}
\begin{figure}
\centering \includegraphics[width=0.55\textwidth]{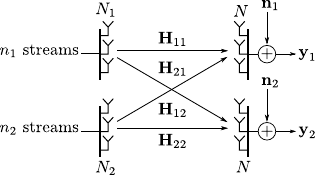}
\caption{Interference channel from two transmitters with $N_k$ ($k=1,2$) antennas, respectively, to two receivers with $N$ antennas each. Each transmitter sends $n_k$  independent data streams to its respective receiver.\label{fig:IC}}
\end{figure}

Consider a MIMO interference channel consisting of two transmitter-receiver pairs as depicted in Fig.~\ref{fig:IC}. The received signal vectors $\yv_1,\yv_2\in\CC^N$ are respectively given as
\begin{align*}
 \yv_1 & = \Hm_{11}\Wm_1\xv_1 + \Hm_{12}\Wm_2\xv_2 + \nv_1\\
 \yv_2 & = \Hm_{21}\Wm_1\xv_1 + \Hm_{22}\Wm_2\xv_2 + \nv_2
\end{align*}
where $\Hm_{qk}\in\CC^{N\times N_k}$, $\Wm_k\in\CC^{N_k\times N_k}$, $\xv_k\sim\Cc\Nc(\zerov,\Pm_k)$, $\Pm_k\in\RR_+^{N_k\times N_k}$ satisfying $\frac1{N_k}\trace\Pm_k=1$, and $\nv_k\sim\Cc\Nc(\zerov,\rho\Id_N)$, for $q,k\in\{1,2\}$. Assuming that the receivers are aware of both precoding matrices and their respective channels but treat the interfering transmission as noise, the normalized mutual information between $\xv_1$ and $\yv_1$, and $\xv_2$ and $\yv_2$ is respectively given as
\begin{align*}
 I_1(\rho) & = \frac1N\log\det\LB\Id_N + \frac1\rho\sum_{k=1}^2\Hm_{1k}\Wm_k\Pm_k\Wm_k\htp\Hm_{1k}\htp\RB-\frac1N\log\det\LB\Id_N+\frac1\rho\Hm_{12}\Wm_2\Pm_2\Wm_2\htp\Hm_{12}\htp\RB\\
I_2(\rho) & = \frac1N\log\det\LB\Id_N + \frac1\rho\sum_{k=1}^2\Hm_{2k}\Wm_k\Pm_k\Wm_k\htp\Hm_{2k}\htp\RB-\frac1N\log\det\LB\Id_N+\frac1\rho\Hm_{21}\Wm_1\Pm_1\Wm_1\htp\Hm_{21}\htp\RB.
\end{align*}
We adopt the same channel model as in Section~\ref{sec:MAC}, where the channel matrices $\Hm_{ik}$ are given as
\begin{align*}
 \Hm_{qk} = \Rm_{qk}^{\frac12}\Um_{qk}\Tm_k^{\frac12}
\end{align*}
where $\Um_{qk}\in\CC^{N\times N_k}$ have independent $\Cc\Nc(0,1/N)$ entries and $\Tm_k$ and $\Rm_{qk}$ are calculated according to \eqref{eq:macchnmodel}. We assume that no channel state information is available at the transmitters, so that the matrices $\Pm_k$ are simply used to determine the number of independently transmitted streams: 
\begin{align*}
\Pm_k=\frac{N_k}{n_k}\diag\LB\underbrace{1,\dots,1}_{n_k},\underbrace{0,\dots,0}_{N_k-n_k}\RB. 
\end{align*}
We will now apply the previously derived results to find the optimal number of streams $(n_1^*,n_2^*)$ maximizing the normalized ergodic sum-rate of the interference channel above. That is, we seek to find
\begin{align*}
 (n_1^*,n_2^*)\ & =\ \max_{n_1,n_2} \mathbb{E}\LSB  I_1(\rho) +  I_2(\rho)\RSB\\
&\quad\text{s.t. } 1\le n_1 \le N_1,\ 1\le  n_2 \le N_2
\end{align*}
where the expectation is with respect to both channel and precoding matrices. Due to the complexity of the random matrix model, this optimization problem appears intractable by exact analysis. At the same time, any solution based on an exhaustive search in combination with Monte Carlo simulations becomes quickly prohibitive for large $N_1,N_2$, since $N_1\times N_2$ possible combinations need to be tested. Relying on Theorem~\ref{th:mutinf}, we can calculate an approximation of $\mathbb{E}\LSB  I_1(\rho) +  I_2(\rho)\RSB$ to find an approximate solution which becomes asymptotically exact as $N_1$ and $N_2$ grow large. Thus, we determine $(\bar{n}_1^*,\bar{n}_2^*)$ as the solution to
\begin{align*}
 (\bar{n}_1^*,\bar{n}_2^*)\ & =\ \max_{n_1,n_2}  \bar{I}_1(\rho) +  \bar{I}_2(\rho)\\
&\quad\text{s.t. } 1\le n_1 \le N_1,\ 1\le  n_2 \le N_2
\end{align*}
where $\bar{I}_1(\rho),\bar{I}_2(\rho)$ are calculated based on a direct application of Theorem~\ref{th:mutinf} to each of the two log-det terms in $I_1(\rho)$ and $I_2(\rho)$, respectively. The optimal values $(\bar{n}_1^*,\bar{n}_2^*)$ are then found by an exhaustive search over all possible combinations. Although we still need to compute $N_1\times N_2$ values, this is computationally much cheaper than Monte Carlo simulations.

\begin{figure}
\centering
\includegraphics[width=0.67\textwidth]{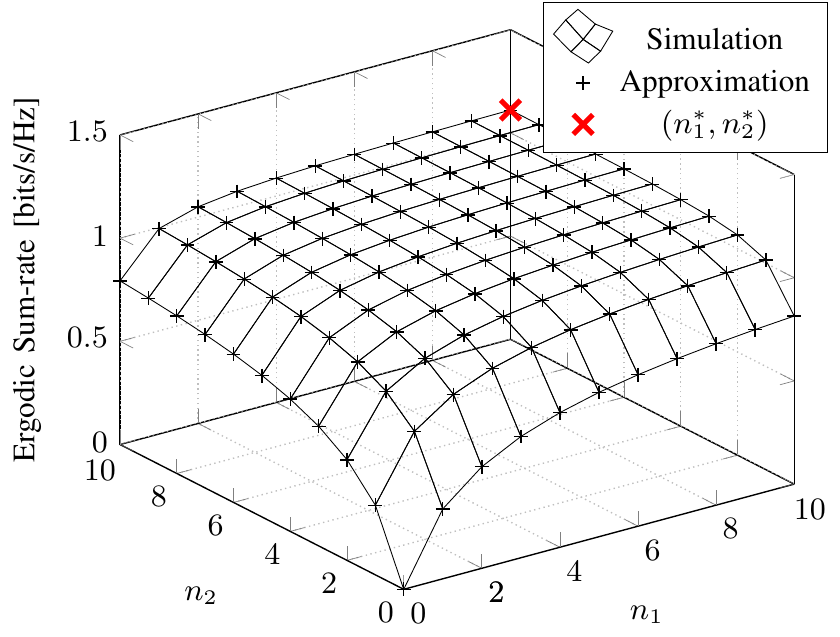}
\caption{Sum-rate versus number of transmitted data-streams $(n_1,n_2)$ for $\text{SNR}=0\,\text{dB}$ and all other parameters as provided in Table~\ref{tab:paramIC}. Solid lines correspond to simulation results, markers to the deterministic approximation by Theorem~\ref{th:mutinf}. As expected, both transmitters should send the maximum number of independent streams.\label{fig:IC_SNR0}}
\end{figure}

\begin{figure}
\centering
\includegraphics[width=0.67\textwidth]{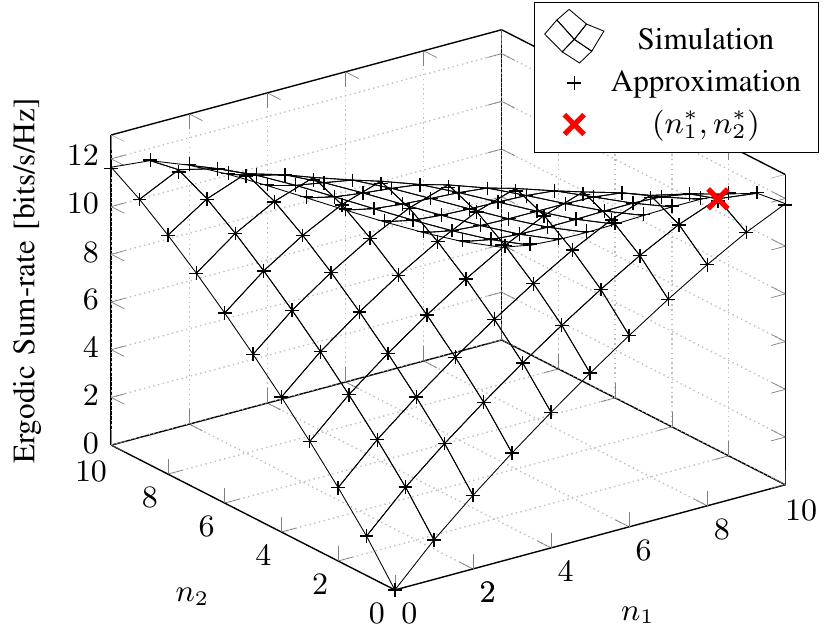}
\caption{Sum-rate versus number of transmitted data-streams $(n_1,n_2)$ for $\text{SNR}=40\,\text{dB}$ and all other parameters as provided in Table~\ref{tab:paramIC}. Solid lines correspond to simulation results, markers to the deterministic approximation by Theorem~\ref{th:mutinf}. As co-channel interference is dominant there is a clear gain of limiting the number of transmitted streams.\label{fig:IC_SNR40}}
\end{figure}

Fig.~\ref{fig:IC_SNR0} and \ref{fig:IC_SNR40} show the average normalized sum-rate $\mathbb{E}\LSB  I_1(\rho) +  I_2(\rho)\RSB$ and the deterministic approximation by Theorem~\ref{th:mutinf} $ \bar{I}_1(\rho) +  \bar{I}_2(\rho)$ as a function of $(n_1,n_2)$ for the simulation parameters as given in Table~\ref{tab:paramIC}. We have assumed $\text{SNR}=0\,\text{dB}$ and $\text{SNR}=40\,\text{dB}$ in Fig.~\ref{fig:IC_SNR0} and \ref{fig:IC_SNR40}, respectively. In both figures, the solid grid represents simulation results and the markers the deterministic approximations. Surprisingly, we observe an almost perfect overlap between both results for all values of $(n_1,n_2)$. The optimal values $(n_1^*,n_2^*)$ and $(\bar{n}_1^*,\bar{n}_2^*)$ coincide for both values of $\text{SNR}$ and are indicated by large crosses in both figures.
At low SNR, both transmitters should send as many independent streams as transmit antennas, i.e., $n_1=n_2=10$. At high SNR, one transmitter should use only a single stream ($n_1=1$) and the other transmitter $n_2=N-1=9$ streams. These results are in line with the observations of \cite{blum2003}.

Obviously, the last result is highly unfair and better solutions can be achieved by using different objective functions, such as weighted sum-rate maximization. Also optimal stream-control with MMSE decoding could be carried out in a similar manner. Although we would still need to perform and exhaustive search over all possible combinations of $n_1,n_2$, the computations are significantly faster than simulation-based approaches. The development of more intelligent algorithms to determine $(\bar{n}^*_1,\bar{n}^*_2)$ is outside the scope of this paper and left to future work. The extensions to more than two transmitter-receiver pairs are straightforward.

\begin{table}
\renewcommand{\arraystretch}{1.3}
\caption{Simulation parameters for Fig.~\ref{fig:IC_SNR0} and \ref{fig:IC_SNR40}: $N=10$, $d^{r,k}_{ij}=4\lambda(i-j)$, $d^{t,k}_{ij}=4\lambda(i-j)$}
\label{tab:paramIC}
\centering
\begin{tabular}{c|ccccc}
$(q,k)$ & $N_k$ & $\theta^{t,k}_\text{min}$ & $\theta^{t,k}_\text{max}$ & $\theta^{r,q,k}_\text{min}$ & $\theta^{r,q,k}_\text{max}$  \\
\hline\hline
(1,1) & 10 & $0$      & $\pi/2$ & $-\pi/4$ & $0$     \\\hline
(1,2) & 10 & $-\pi/2$ & $0$     & $0$      & $\pi/4$ \\\hline
(2,1) & 10 & $0$      & $\pi/2$ & $-\pi/3$ & $0$     \\\hline
(2,2) & 10 & $-\pi/2$ & $0$     & $0$      & $\pi/3$ \\\hline
\end{tabular}
\end{table}

\clearpage
\section{Conclusion}\label{sec:con}
In this correspondence, we have studied a general channel model for randomly isometric precoded MIMO systems over correlated fading channels which finds useful applications in the context of multiple access and interference channels. For the MAC, we have derived deterministic approximations of the normalized (ergodic) mutual information, the (ergodic) sum-rate with single-stream MMSE decoding as well as the SINR of the MMSE receiver, which are almost surely asymptotically tight. Moreover, we have provided an asymptotically optimal power allocation algorithm under individual and sum-power constraints. Our results were then used for optimal stream control in interference channels. Numerical results show that the asymptotic results provide very tight approximations for systems with realistic dimensions.

\clearpage
\appendices
\section{Proofs}\label{app:proofs}
\begin{IEEEproof}[Proof of Theorem~\ref{th:fundeq} (Fundamental equations)]
 It was shown in \cite[Eq. (27)]{couillet10b} that, for any fixed $g_k(\rho)\ge0$, the following equation in $\bar{g}_k(\rho)$:
\begin{align*}
 \bar{g}_k(\rho) = \frac1N\trace\Pm_k\Big( g_k(\rho)\Pm_k + \LSB\bar{c}_k-g_k(\rho)\bar{g}_k(\rho)\RSB\Id_{n_k}\Big)^{-1}
\end{align*}
has a unique solution, satisfying $0\le \bar{g}_k(\rho)< c_k\bar{c}_k/g_k(\rho)$. Thus, $\bar{g}_k(\rho)$ is uniquely determined by $g_k(\rho)$. 
Consider now the following functions for $k\in\{1,\dots,K\}$ and $\rho>0$:
\begin{align*}
 h_k(x_1,\dots,x_K) \mapsto \frac1N\sum_{j=1}^{N_k}\frac{\delta_{kj}(\rho)}{1+\bar{g}_k\delta_{kj}(\rho)}
\end{align*}
where $\bar{g}_k\in[0,c_k\bar{c}_k/x_k)$ and $\delta_{kj}(\rho)\ge 0$ are the unique solutions to the following fixed-point equations:
\begin{align}\label{eq:bare}
 \bar{g}_k & = \frac1N\trace\Pm_k\Big( x_k\Pm_k + \LSB\bar{c}_k-x_k\bar{g}_k\RSB\Id_{n_k}\Big)^{-1}\\\label{eq:T}
\delta_{kj}(\rho) &=\frac1N\trace\Rm_{kj} \LB\frac{1}{N}\sum_{k=1}^K\sum_{j=1}^{N_k}\frac{\bar{g}_k\Rm_{k,j}}{1+\bar{g}_k\delta_{kj}(\rho)}+\rho\Id_N\RB^{-1}.
\end{align}
Similar to \cite[Proof of Theorem 1, Step 2]{couillet10b} it is now sufficient to prove that the $K$-variate function $\hv: (x_1,\dots,x_K)\mapsto (h_1,\ldots,h_K)$ is a \emph{standard function} \cite{yates95}:

\vspace{10pt}\begin{definition}
A function $\hv(x_1,\ldots,x_K)\in \RR^K$ is \emph{standard} if it fulfills the following conditions:
\begin{enumerate}
	\item {\it Positivity:} if $x_1,\ldots,x_K>0$, then for all $k$, $h_k(x_1,\ldots,x_K)>0$.
	\item {\it Monotonicity:} if $x_1>x_1',\ldots,x_K>x_K'$, then for all $k$, $h_k(x_1,\ldots,x_K)>h_k(x_1',\ldots,x_K')$.
	\item {\it Scalability:} for all $\alpha>1$ and $k$, $\alpha h_k(x_1,\ldots,x_K)>h_k(\alpha x_1,\ldots,\alpha x_K)$.
\end{enumerate}
\end{definition}\vspace{10pt}
This guarantees by \cite[Theorem 2]{yates95} that a standard fixed-point algorithm that consists of setting
\begin{align*}
 x_k^{(t+1)} = h_k(x_1^{(t)},\dots,x_K^{(t)}),\qquad k=1,\dots,K
\end{align*}
for $t\ge0$ and for any set of initial values $x_1^{(0)},\dots,x_K^{(0)}>0$, converges to the unique jointly positive solution of the system of $K$ equations
\begin{align*}
 x_k = h_k(x_1,\dots,x_K),\qquad k=1,\dots,K.
\end{align*}

Showing positivity is straightforward: For $\rho>0$, we have $\delta_{kj}(\rho)> 0$ by Theorem~\ref{th:detequcorr} in Appendix~\ref{sec:appendixB} and $\bar{g}_k\ge0$ by its definition. Thus, $h_k(x_1,\dots,x_K)> 0$ for all $x_1,\ldots,x_K>0$.

To prove monotonicity of $h_k(x_1,\dots,x_K)$, we need the following results:

\vspace{10pt}\begin{lemma}[{\cite[Eq. (26)]{couillet10b}}]\label{lem:gk}
 Let $x_k > x_k'$, and consider $\bar{g}_k$ and $\bar{g}_k'$ the corresponding solutions to \eqref{eq:bare}. Then,
\begin{align*}
 \text{(i)}\ \ \bar{g}_k<\bar{g}_k' \qquad\qquad \text{(ii)}\ \  x_k\bar{g}_k>x_k'\bar{g}_k'.
\end{align*}
\end{lemma}\vspace{10pt}

\vspace{10pt}\begin{lemma}\label{lem:gkbar}
 Let $\rho>0$ and assume $\bar{g}_k > \bar{g}_k'$. Consider $\delta_{kj}(\rho)$ and $\delta_{kj}'(\rho)$ as the unique solutions to \eqref{eq:T} for $\bar{g}_k$ and $\bar{g}_k'$, respectively. Then,
\begin{align*}
 \text{(i)}\ \ \delta_{kj}(\rho)<\delta'_{kj}(\rho) \qquad\qquad \text{(ii)}\ \ \bar{g}_k\delta_{kj}(\rho)>\bar{g}_k'\delta'_{kj}(\rho).
\end{align*}
\end{lemma}\vspace{10pt}

\begin{IEEEproof}
The proof is based on the consideration of an extended version of the random matrix model assumed in Theorem~\ref{th:detequcorr}. Let us consider the following random matrices $\Hm_k^L\in\CC^{LN\times LN_k}$, given as
\begin{align*}
 \Hm_k^L = \frac{1}{\sqrt{LN}}\LSB\LB\Rm^L_{k1}\RB^{\frac12}\Um^L_{k1},\dots,\LB\Rm^L_{kN_k}\RB^{\frac12}\Um^L_{kN_k}\RSB
\end{align*}
where $\Rm^L_{kj}=\diag(\Rm_{kj},\dots,\Rm_{kj})\in\CC^{LN\times LN}$ are block-diagonal matrices consisting of $L$ copies of the matrix $\Rm_{kj}$ and $\Um^L_{kj}\in\CC^{LN\times L}$ are random matrices composed of i.i.d.\@ entries with zero mean, unit variance and finite moment of order $4+\epsilon$, for some $\epsilon>0$. 
We define the following matrices which will be of repeated use in the sequel:
\begin{align*}
\tilde{\Bm}^L&=\sum_{k=1}^K\bar{g}_k\Hm_k^L\LB\Hm_k^L\RB\htp, \qquad \tilde{\Bm'}^L=\bar{g}_k'\Hm_k^L\LB\Hm_k^L\RB\htp + \sum_{l=1,l\ne k}^K\bar{g}_l\Hm_l^L\LB\Hm_l^L\RB\htp\\
\Qm&=\LB\tilde{\Bm}^L +\rho\Id_{NL}\RB^{-1}, \qquad \Qm'=\LB\tilde{\Bm'}^L +\rho\Id_{NL}\RB^{-1}.
\end{align*}
One can verify from Theorem~\ref{th:detequcorr} that for any fixed $N,N_1,\dots,N_K$, the following limit holds:
\begin{align*}
 \frac{1}{LN}\trace\Rm_{kj}^L\LB\tilde{\Bm}^L +\rho\Id_{NK}\RB^{-1}  \xrightarrow[L\to\infty]{\text{a.s}}\delta_{kj}(\rho).
\end{align*}
Thus, any properties of the random quantities on the left-hand side of the previous equation also hold for the deterministic quantities $\delta_{kj}(\rho)$. We will exploit this fact for the termination of the proof.
Notice that the matrices $\tilde{\Bm}^L$ and $ \tilde{\Bm'}^L$ differ only by $\bar{g}_k$. This is sufficient since the case $\bar{g}_l>\bar{g}_l'$ for $l\in\{1,\dots,K\}$ follows by simple iteration of the case $\bar{g}_l=\bar{g}_l'$ for $l\neq k$ and $\bar{g}_k>\bar{g}_k'$.

To proof (i), it is now sufficient to show that, for any $L$,
\begin{align*}
 \frac1N\trace\Rm^L_{k,j}\LB\Qm-\Qm'\RB <0.
\end{align*}
By Lemma~\ref{lem:traceinequ}, this is equivalent to proving $\LB\Qm\RB^{-1}-\LB\Qm'\RB^{-1}\succ0$, which is straightforward since
\begin{align*}
 \LB\Qm\RB^{-1}-\LB\Qm'\RB^{-1} &= \tilde{\Bm}^L - \tilde{\Bm'}^L = (\bar{g}_k-\bar{g}_k')\Hm_k^L\LB\Hm_k^L\RB\htp \succ 0.
\end{align*}
Thus,
\begin{align*}
 \frac1N\trace\Rm^L_{k,j}\LB\Qm-\Qm'\RB \xrightarrow[L\to\infty]{\text{a.s}}\delta_{kj}(\rho)-\delta'_{kj}(\rho)< 0
\end{align*}
since $\delta_{kj}(\rho)$ and $\delta'_{kj}(\rho)$ do not depend on $L$.

For (ii), we need to show that
\begin{align*}
 \bar{g}_k\frac{1}{LN}\trace\Rm_{kj}^L\Qm - \bar{g}_k'\frac{1}{LN}\trace\Rm_{kj}^L\Qm'>0.
\end{align*}
Similarly to the previous part of the proof, it is sufficient to show that $\LB\bar{g}_k\Qm\RB^{-1}-\LB\bar{g}_k'\Qm'\RB^{-1}\prec0$. Hence,
\begin{align*}
 \LB\bar{g}_k\Qm\RB^{-1}-\LB\bar{g}_k'\Qm'\RB^{-1} &= \frac{1}{\bar{g}_k}\LB\tilde{\Bm}^L+\rho\Id_{NL}\RB - \frac{1}{\bar{g}'_k}\LB\tilde{\Bm'}^L+\rho\Id_{NL}\RB\\
&= \rho\LB\frac{1}{\bar{g}_k}-\frac{1}{\bar{g}'_k}\RB\Id_{NL} + \LB\frac{1}{\bar{g}_k}-\frac{1}{\bar{g}'_k}\RB\sum_{l=1,l\ne k}^K\bar{g}_l\Hm_l^L\LB\Hm_l^L\RB\htp\\
&\prec 0
\end{align*}
since $\rho>0$, $\bar{g}_k>\bar{g}_k'$ and $\bar{g}_l\ge0$ for all $l$.
\end{IEEEproof}\vspace{10pt}

Consider now $(x_1,\dots,x_K)$ and $(x_1',\dots,x_K')$, such that $x_k>x_k'\ \forall k$, and denote by $(\bar{g}_1,\dots,\bar{g}_K$) and $(\bar{g}'_1,\dots,\bar{g}'_K)$ the corresponding solutions to \eqref{eq:bare}. Denote by $\delta_{kj}(\rho)$ and $\delta'_{kj}(\rho)$ the unique solutions to \eqref{eq:T} for $(\bar{g}_1,\dots,\bar{g}_K)$ and $(\bar{g}'_1,\dots,\bar{g}'_K)$, respectively.
It follows from Lemma~\ref{lem:gk}, that $\bar{g}_k<\bar{g}_k'\ \forall k$. Lemma~\ref{lem:gkbar} now implies that $\delta_{kj}(\rho)>\delta'_{kj}(\rho)$ and $\bar{g}_k\delta_{kj}(\rho)<\bar{g}_k'\delta_{kj}'(\rho)$. Combining these results yields
\begin{align*}
 h_k(x_1,\dots,x_K) = \frac1N\sum_{j=1}^{N_k}\frac{\delta_{kj}(\rho)}{1+\bar{g}_k\delta_{kj}(\rho)} > \frac1N\sum_{j=1}^{N_k}\frac{\delta'_{kj}(\rho)}{1+\bar{g}_k'\delta'_{kj}(\rho)} =h_k(x_1',\dots,x_K')
\end{align*}
which proves monotonicity.

To proof scalability, let $\alpha>1$, and consider the following difference:
\begin{align*}
 \alpha h_k(x_1,\dots,x_K) -  h_k(\alpha x_1,\dots,\alpha x_K) &= \frac1N \sum_{j=1}^{N_k} \frac{\alpha \delta_{kj}(\rho)}{1+\bar{g}_k\delta_{kj}(\rho)}-\frac{\delta^{(\alpha)}_{kj}(\rho)}{1+\bar{g}_k^{(\alpha)}\delta_{kj}^{(\alpha)}(\rho)}\\
&= \frac1N \sum_{i=1}^{N_k} \frac{\LSB\alpha \delta_{kj}(\rho) - \delta_{kj}^{(\alpha)}(\rho)\RSB+ \delta_{kj}(\rho)  \delta_{kj}^{(\alpha)}(\rho)\LSB\alpha\bar{g}_k^{(\alpha)} - \bar{g}_k\RSB}{\LSB1+\bar{g}_k  \delta_{kj}(\rho)\RSB\LSB1+\bar{g}_k^{(\alpha)} \delta_{kj}^{(\alpha)}(\rho)\RSB}
 \end{align*}
where we have denoted by $\bar{g}_k^{(\alpha)}$ the solution to \eqref{eq:bare} with $x_k$ replace by $\alpha x_k$  and by $\delta_{kj}^{(\alpha)}(\rho)$ the solution to \eqref{eq:T} for  $\bar{g}_k^{(\alpha)}$. We have from Lemma~\ref{lem:gk}~(i) that $\bar{g}_k^{(\alpha)}<\bar{g}_k$ and from Lemma~\ref{lem:gk}~(ii) that
\begin{align}\label{eq:inqu1}
 \alpha x_k \bar{g}_k^{(\alpha)} > x_k \bar{g}_k \Longleftrightarrow \alpha \bar{g}_k^{(\alpha)} -\bar{g}_k>0.
\end{align}
It remains now to show that also $\alpha \delta_{kj}(\rho)-\delta_{kj}^{(\alpha)}(\rho)>0$. 
To this end, consider the following difference:
\begin{align}\nonumber
 \alpha \delta_{kj}(\rho)-\delta_{kj}^{(\alpha)}(\rho)& = \frac1N\trace\Rm_{kj}\LB\alpha\Tm(\rho)-\Tm^{(\alpha)}(\rho)\RB
\end{align}
where 
\begin{align*}
 \Tm(\rho)&=\LB\frac{1}{N}\sum_{k=1}^K\sum_{j=1}^{N_k}\frac{\bar{g}_k\Rm_{k,j}}{1+\bar{g}_k\delta_{kj}(\rho)}+\rho\Id_N\RB^{-1}\\
\Tm^{(\alpha)}(\rho)&=\LB\frac{1}{N}\sum_{k=1}^K\sum_{j=1}^{N_k}\frac{\bar{g}_k^{(\alpha)}\Rm_{k,j}}{1+\bar{g}_k^{(\alpha)}\delta_{kj}^{(\alpha)}(\rho)}+\rho\Id_N\RB^{-1}.
\end{align*}
By Lemma~\ref{lem:traceinequ}, it is now sufficient to show that $\LB\Tm^{(\alpha)}(z)\RB^{-1}\succ\LB\alpha\Tm(z)\RB^{-1}$. Write therefore
\begin{align*}
&\ \LB\Tm^{(\alpha)}(\rho)\RB^{-1}- \LB\alpha\Tm(\rho)\RB^{-1}\\=\ &\ \rho\LB1-\frac1\alpha\RB\Id_N +\frac1N\sum_{k=1}^K\sum_{j=1}^{N_k}\frac{\LSB\alpha\bar{g}_k^{(\alpha)}-\bar{g}_k\RSB+\bar{g}_k^{(\alpha)}\bar{g}_k\LSB\alpha \delta_{kj}(\rho)-\delta_{kj}^{(\alpha)}(\rho)\RSB}{\alpha\LSB1+\bar{g}_k \delta_{kj}(\rho)\RSB\LSB1+\bar{g}_k^{(\alpha)}\delta_{kj}^{(\alpha)}(\rho)\RSB}\Rm_{kj}.
\end{align*}
The first summand is positive definite since $\rho>0$ and $\alpha>1$. All other terms are also positive definite since $\alpha\bar{g}_k^{(\alpha)}-\bar{g}_k>0$ from \eqref{eq:inqu1} and $\alpha\bar{g}_k^{(\alpha)}\bar{g}_k\delta_{kj}(\rho)>\bar{g}_k\bar{g}_k^{(\alpha)}\delta_{kj}^{(\alpha)}(\rho)$, since $\alpha\bar{g}_k^{(\alpha)}>\bar{g}_k$ and $\bar{g}_k\delta_{kj}(\rho)>\bar{g}_k^{(\alpha)}\delta_{kj}^{(\alpha)}(\rho)$ by Lemma~\ref{lem:gkbar}~(ii) and Lemma~\ref{lem:gk}~(i). Since the sum of positive definite matrices is also positive definite, we have $\alpha \delta_{kj}(\rho)-\delta_{kj}^{(\alpha)}(\rho)>0$. This terminates the proof of scalability.

Thus, we have shown $\hv: (x_1,\dots,x_K)\mapsto (h_1,\ldots,h_K)$ to be a standard function.
Moreover, from \cite[Remark to Theorem 1]{couillet10b} and Theorem~\ref{th:detequcorr}, we have the following algorithm to compute $\bar{g}_k$ and $\delta_{kj}(\rho)$:
\begin{align*}
 \bar{g}_k = \lim_{t\to\infty}  \bar{g}_k^{(t)}, \qquad\delta_{kj}(\rho)= \lim_{t\to\infty}  \delta_{kj}^{(t)}(\rho)
\end{align*}
where
\begin{align*}
 \bar{g}_k^{(t)} & = \frac1N\trace\Pm_k\Big( x_k\Pm_k + \LSB\bar{c}_k-x_k\bar{g}_k^{(t-1)}\RSB\Id_{n_k}\Big)^{-1}\\
 \delta_{kj}^{(t)}(\rho) &= \frac1N\trace\Rm_{kj} \LB\frac{1}{N}\sum_{k=1}^K\sum_{j=1}^{N_k}\frac{\bar{g}_k\Rm_{k,j}}{1+\bar{g}_k\delta_{kj}^{(t-1)}(\rho)}+\rho\Id_N\RB^{-1}
\end{align*}
and $\bar{g}_k^{(0)}$ can take any value in $[0,c_k\bar{c}_k/x_k)$ and $\delta_{kj}^{(0)}(\rho)=1/\rho$ for all $k,j$.
\end{IEEEproof}

\clearpage
\begin{IEEEproof}[Proof of Theorem 2 (Mutual information)]
 We begin by proving the following result: 
\begin{align}\label{eq:conv1}
 \max_k|\bar{e}_k(\rho) - \bar{g}_k(\rho)| &\xrightarrow[N\to\infty]{\text{a.s.}}0\\\label{eq:conv2}
 \max_k|e_k(\rho) - g_k(\rho)| &\xrightarrow[N\to\infty]{\text{a.s.}}0
\end{align}
where $\bar{e}_k(\rho)$, $e_k(\rho)$ are defined in Theorem~\ref{th:fundequdet} and $\bar{g}_k(\rho)$, $g_k(\rho)$ are defined in Theorem~\ref{th:fundeq}, assuming that the matrices $\Hm_k$ are random and modeled as described in \eqref{eq:channelmodel}.
For notational simplicity, we will drop from now on the dependence on $\rho$.  For any given family $(\bar{f}_{N,1},\dots,\bar{f}_{N,K})$, $N=1,2,\dots$, of bounded real numbers,we have from standard lemmas of random matrix theory:\footnote{Let $a_N$ and $b_N$ denote a pair of infinite sequences of random variables. We write $a_N\asymp b_N$, iff $a_N-b_N\xrightarrow[]{\text{a.s.}}0$ for $N\to\infty$.}
\begin{align}\nonumber
 f_{N,k} &\defines \frac1N\trace \Hm_k\Hm_k\htp\LB\sum_{i=1}^K \bar{f}_{N,i} \Hm_i\Hm_i\htp +\rho\Id_N \RB^{-1}\\\nonumber
&= \frac1N\sum_{j=1}^{N_k} \hv_{kj}\htp\LB\sum_{i=1}^K \bar{f}_{N,i} \Hm_i\Hm_i\htp +\rho\Id_N \RB^{-1} \hv_{kj}\\\nonumber
&\overset{\text{(a)}}{=}\frac1N\sum_{j=1}^{N_k} \frac{\hv_{kj}\htp\LB\sum_{i=1}^K \bar{f}_{N,i} \Hm_i\Hm_i\htp - \bar{f}_{N,k}\hv_{kj}\hv_{kj}\htp +\rho\Id_N \RB^{-1} \hv_{kj}}{1+\bar{f}_{N,k}\hv_{kj}\htp\LB\sum_{i=1}^K \bar{f}_{N,i} \Hm_i\Hm_i\htp- \bar{f}_{N,k}\hv_{kj}\hv_{kj}\htp +\rho\Id_N \RB^{-1} \hv_{kj}}\\\nonumber
&\overset{\text{(b)}}{\asymp} \frac1N\sum_{j=1}^{N_k} \frac{\frac1N\trace\Rm_{kj}\LB\sum_{i=1}^K \bar{f}_{N,i} \Hm_i\Hm_i\htp - \bar{f}_{N,k}\hv_{kj}\hv_{kj}\htp +\rho\Id_N \RB^{-1} }{1+\bar{f}_{N,k}\frac1N\trace\Rm_{kj}\LB\sum_{i=1}^K \bar{f}_{N,i} \Hm_i\Hm_i\htp- \bar{f}_{N,k}\hv_{kj}\hv_{kj}\htp +\rho\Id_N \RB^{-1}}\\\nonumber
&\overset{\text{(c)}}{\asymp} \frac1N\sum_{j=1}^{N_k} \frac{\frac1N\trace\Rm_{kj}\LB\sum_{i=1}^K \bar{f}_{N,i} \Hm_i\Hm_i\htp +\rho\Id_N \RB^{-1} }{1+\bar{f}_{N,k}\frac1N\trace\Rm_{kj}\LB\sum_{i=1}^K \bar{f}_{N,i} \Hm_i\Hm_i\htp +\rho\Id_N \RB^{-1}}\\\label{eq:ekconv}
&\overset{\text{(d)}}{\asymp} \frac1N\sum_{j=1}^{N_k} \frac{\frac1N\trace\Rm_{kj}\bar{\Tm} }{1+\bar{f}_{N,k}\frac1N\trace\Rm_{kj}\bar{\Tm}}
\end{align}
where (a) follows from Lemma~\ref{lem:inversion}, (b) follows from Lemma~\ref{lem:trace} and Lemma~\ref{lem:convergence_ratios}, (c) is due to Lemma~\ref{lem:rank1perturbation} and (d) follows from a direct application of Theorem~\ref{th:detequcorr}, where we have defined
\begin{align*}
 \bar{\Tm} = \LB\frac{1}{N}\sum_{k=1}^K\sum_{j=1}^{N_k}\frac{\bar{f}_{N,k}\Rm_{kj}}{1+\bar{f}_{N,k}\frac1N\trace\Rm_{kj}\bar{\Tm}}+\rho\Id_N\RB^{-1}.
\end{align*}
Hence, in particular for $\bar{f}_{N,k}=\bar{e}_k$ and $f_{N,k}=e_k$ as defined in Theorem~\ref{th:fundequdet}, we can write
\begin{align*}
  e_k & =\frac1N\trace \Hm_k\Hm_k\htp\LB\sum_{i=1}^K \bar{e}_i \Hm_i\Hm_i\htp +\rho\Id_N \RB^{-1}= \frac1N\sum_{j=1}^{N_k} \frac{\frac1N\trace\Rm_{kj}\bar{\Tm} }{1+\bar{e}_k\frac1N\trace\Rm_{kj}\bar{\Tm}} + \epsilon_{N,k}
\end{align*}
for some sequences of reals $\epsilon_{N,k}$, satisfying $\epsilon_{N,k}\xrightarrow[N\to\infty]{\text{a.s.}}0$.

Recall now the following definitions for $k=1,\dots,K$:
\begin{align*}
 e_k & = \frac1N\sum_{j=1}^{N_k} \frac{\frac1N\trace\Rm_{kj}\bar{\Tm} }{1+\bar{e}_k\frac1N\trace\Rm_{kj}\bar{\Tm}} + \epsilon_{N,k}\\
g_k & = \frac1N\sum_{j=1}^{N_k} \frac{\frac1N\trace\Rm_{kj}\Tm }{1+\bar{g}_k\frac1N\trace\Rm_{kj}\Tm} \\
\bar{e}_k & = \frac1N\sum_{j=1}^{n_k}\frac{p_{kj}}{\bar{c}_k - e_k\bar{e}_k + e_k p_{kj}},\qquad 0\le\bar{e}_k<c_k\bar{c}_k/e_k\\
\bar{g}_k & = \frac1N\sum_{j=1}^{n_k}\frac{p_{kj}}{\bar{c}_k - g_k\bar{g}_k + g_k p_{kj}},\qquad 0\le\bar{g}_k<c_k\bar{c}_k/g_k
\end{align*}
where 
\begin{align*}
 \bar{\Tm} &= \LB\frac{1}{N}\sum_{k=1}^K\sum_{j=1}^{N_k}\frac{\bar{e}_{k}\Rm_{kj}}{1+\bar{f}_{N,k}\frac1N\trace\Rm_{kj}\bar{\Tm}}+\rho\Id_N\RB^{-1}\\
\Tm &= \LB\frac{1}{N}\sum_{k=1}^K\sum_{j=1}^{N_k}\frac{\bar{g}_k\Rm_{kj}}{1+\bar{g}_k\frac1N\trace\Rm_{kj}\Tm}+\rho\Id_N\RB^{-1}.
\end{align*}
\subsection{Case: $\lim\sup c_k<1$}
We will first assume that $\lim\sup c_k<1$ for all $k$. The case $\lim\sup c_k=1$ will be treated separately in the subsequent section.
Denote $P = \max_k\{\lim\sup\lVert\Pm_k\rVert\}$, $R = \max_m\{\lim\sup\lVert\tilde{\Rm}_{m}\rVert\}$, $c_+ = \max_k\{\lim\sup c_k\}$ and $\bar{c}_- = \min_k\{\lim\inf\bar{c}_k\}$, $\bar{c}_+ = \max_k\{\lim\sup\bar{c}_k\}$. 
Since we are interested in the asymptotic limit $N\to\infty$, we assume from the beginning that $N$ is sufficiently large, so that the following inequalities hold for all $k$:
\begin{align*}
 c_k \le c_+,\quad \bar{c}_- \le \bar{c}_k \le \bar{c}_+,\quad \lVert\Pm_k\rVert\le P,\quad \lVert\Rm_{kj}\rVert\le R.
\end{align*}
We then have the following properties:
\begin{align}\label{eq:inequcase1}
 \bar{e}_k \le \frac{P}{(1-c_+)\bar{c}_-},\quad \bar{g}_k \le \frac{P}{(1-c_+)\bar{c}_-},\quad g_k\bar{g}_k < c_+\bar{c}_+, \quad e_k\bar{e}_k < c_+\bar{c}_+.
\end{align}
For notational simplicity, we define the following quantities:
\begin{align*}
 \alpha = \max_k|e_k-g_k|,\qquad  \bar{\alpha} = \max_k|\bar{e}_k-\bar{g}_k|.
\end{align*}
We will show in the sequel that, almost surely, $\alpha\to 0$ and $ \bar{\alpha}\to 0$ as $N\to\infty$.

Consider first the following difference:
\begin{align*}
 \sup_{k,j}\left|\frac1N\trace\Rm_{kj}\LB\Tm-\bar{\Tm}\RB \right|&= \sup_{k,j}\left|\frac1N\trace\Rm_{kj}\Tm\LB\frac1N\sum_{l=1}^K\sum_{m=1}^{N_l}\frac{\bar{e}_l\Rm_{lm}}{1+\bar{e}_l\frac1N\trace\Rm_{lm}\bar{\Tm}}-\frac{\bar{g}_l\Rm_{lm}}{1+\bar{g}_l\frac1N\trace\Rm_{lm}\bar{\Tm}}\RB\bar{\Tm}\right|\\
&= \sup_{k,j}\left|\frac1N\sum_{l=1}^K\sum_{m=1}^{N_l}\frac{\bar{e}_l-\bar{g}_l + \bar{e}_l\bar{g}_l\LB\frac1N\trace\Rm_{lm}\Tm-\frac1N\trace\Rm_{lm}\bar{\Tm}\RB}{\LB1+\bar{e}_l\frac1N\trace\Rm_{lm}\bar{\Tm}\RB\LB1+\bar{g}_l\frac1N\trace\Rm_{lm}\bar{\Tm}\RB} \frac1N\trace\Rm_{kj}\bar{\Tm}\Rm_{lm}\Tm \right|\\
&\le \frac{R^2}{\rho^2}K\max_k\bar{c}_k\LSB \max_{k}|\bar{e}_k-\bar{g}_k|+\max_k|\bar{e}_k\bar{g}_k|\sup_{k,j}\left|\frac1N\trace\Rm_{kj}\LB\Tm-\bar{\Tm}\RB\right|\RSB\\
&\le \frac{R^2}{\rho^2}K\bar{c}_+\LSB\bar{\alpha}+\frac{P^2}{(1-c_+)^2\bar{c}_-^2}\sup_{k,j}\left|\frac1N\trace\Rm_{kj}\LB\Tm-\bar{\Tm}\RB\right|\RSB
\end{align*}
where the first equality follows from Lemma~\ref{lem:resolvent}. Rearranging the terms yields:
\begin{align}\label{eq:inequ1}
 \sup_{k,j}\left|\frac1N\trace\Rm_{kj}\LB\Tm-\bar{\Tm}\RB\right| \le  \frac{P^2K\bar{c}_+}{\rho^2 - \frac{R^2P^2}{(1-c_+)^2\bar{c}_-^2}}\ \bar{\alpha}
\end{align}
for $\rho>\frac{RP}{(1-c_+)\bar{c}_-} $.

Consider now the term $\alpha=\max_k|e_k-g_k|$:
\begin{align}\nonumber
 \alpha &= \max_k\left|\frac1N\sum_{j=1}^{N_k}\frac{\frac1N\trace\Rm_{kj}\LB\bar{\Tm}-\Tm\RB + (\bar{g}_k-\bar{e}_k)\frac1N\trace\Rm_{kj}\frac1N\trace\Rm_{kj}\bar{\Tm}}{\LB1+\bar{e}_k\frac1N\trace\Rm_{kj}\bar{\Tm}\RB\LB1+\bar{g}_k\frac1N\trace\Rm_{kj}\Tm\RB} +\epsilon_{N,k} \right|\\\nonumber
&\le \bar{c}_+\sup_{kj}\left|\frac1N\trace\Rm_{kj}\LB\Tm-\bar{\Tm}\RB\right| + \bar{c}_+\frac{R^2}{\rho^2}\max_k |\bar{e}_k-\bar{g}_k|+\max_k|\epsilon_{N,k}|\\\nonumber
&\le \frac{P^2K\bar{c}_+^2}{\rho^2 - \frac{R^2P^2}{(1-c_+)^2\bar{c}_-^2}} \bar{\alpha} + \frac{\bar{c}_+R^2}{\rho^2} \bar{\alpha} + \max_k|\epsilon_{N,k}|\\\label{eq:inequalpha}
&= \LSB\frac{P^2K\bar{c}_+^2}{\rho^2 - \frac{R^2P^2}{(1-c_+)^2\bar{c}_-^2}} + \frac{\bar{c}_+R^2}{\rho^2} \RSB \bar{\alpha} + \max_k|\epsilon_{N,k}|
\end{align}
where the last inequality follows from \eqref{eq:inequ1}.
Similarly, we have for $\bar{\alpha}=\max_k|\bar{e}_k-\bar{g}_k|$:
\begin{align*}
 \bar{\alpha} & = \max_k\left|\frac1N\sum_{j=1}^{n_k}p_{kj} \frac{e_k\bar{e}_k-g_k\bar{g}_k+p_{kj}(g_k-e_k)}{(\bar{c}_k-e_k\bar{e}_k +e_k p_{kj})(\bar{c}_k-g_k\bar{g}_k +g_k p_{kj})} \right|\\
&\le \frac1N\sum_{j=1}^{n_k} \frac{p_{kj}^2\max_k|e_k-g_k|}{(1-c_+)^2\bar{c}_-^2} + p_{kj}\frac{\max_k\LSB \bar{e}_k|e_k-g_k\RSB| + \max_k\LSB g_k|\bar{e}_k-\bar{g}_k|\RSB  }{(1-c_+)^2\bar{c}_-^2}\\
&\le\frac{P^2}{(1-c_+)^2\bar{c}_-^2}\LB1 + \frac{1}{(1-c_+)\bar{c}_-}\RB \alpha + \frac{PR\bar{c}_+}{\rho(1-c_+)^2\bar{c}_-^2}\bar{\alpha}.
\end{align*}
Thus, for $\rho\ge\max\left\{ \frac{2PR\bar{c}_+}{(1-c_+)^2\bar{c}_-^2},\frac{RP}{(1-c_+)\bar{c}_-}\right\}$, we have
\begin{align}\label{eq:inequalphabar}
 \bar{\alpha}\le \frac{2P^2}{(1-c_+)^2\bar{c}_-^2}\LB1 + \frac{1}{(1-c_+)\bar{c}_-}\RB \alpha.
\end{align}
Replacing \eqref{eq:inequalphabar} in \eqref{eq:inequalpha} leads to
\begin{align*}
 \alpha &\le\LSB\frac{P^2K\bar{c}_+^2}{\rho^2 - \frac{R^2P^2}{(1-c_+)^2\bar{c}_-^2}} + \frac{\bar{c}_+R^2}{\rho^2} \RSB \frac{2P^2}{(1-c_+)^2\bar{c}_-^2}\LB1 + \frac{1}{(1-c_+)\bar{c}_-}\RB \alpha + \max_k|\epsilon_{N,k}|.
\end{align*}
For $\rho$ sufficiently large, we therefore have
\begin{align*}
 \alpha \le 2 \max_k|\epsilon_{N,k}| \xrightarrow[N\to\infty]{\text{a.s.}} 0 .
\end{align*}
This implies by \eqref{eq:inequalphabar} that also $\bar{\alpha} \xrightarrow[N\to\infty]{\text{a.s.}} 0$ .
Since $g_k,\bar{e}_k,\bar{g}_k$ are uniformly bounded on all closed subsets of $\RR_+$ and $e_k$ is almost surely uniformly bounded on all closed subsets of $\RR_+$, we have from Vitali's convergence theorem \cite{titchmarsh} that the almost sure convergence holds true for all $\rho\in\RR_+$. This terminates the proof for $c_k<1$.

\subsection{Case: $\lim\sup c_k=1$}
It was shown in \cite[Proof of Theorem 1]{couillet10b}, that the following refined inequalities hold for $c_k=1$:
\begin{align*}
 \bar{e}_k \le P,\quad \bar{g}_k \le P.
\end{align*}
Using these inequalities instead of \eqref{eq:inequcase1} in the proof of the case $c_k<1$, one can show that 
$\alpha \xrightarrow[]{\text{a.s.}} 0$ and $\bar{\alpha} \xrightarrow[]{\text{a.s.}} 0$ as $N\to\infty$.

\subsection{Convergence of the mutual information}
Consider now the first term of $\Vc_N(\rho)$ in Theorem~\ref{th:logdetdet}. Due to the convergence of $\bar{e}_k-\bar{g}_k\xrightarrow[]{\text{a.s.}}0$, we have from the continuous mapping theorem \cite[Theorem 2.3]{vdv} that
\begin{align*}
 \frac1N\log\det\LB\Id_N+\frac{1}{\rho}\sum_{k=1}^K\bar{e}_k\Hm_k\Hm_k\htp\RB-\frac1N\log\det\LB\Id_N+\frac{1}{\rho}\sum_{k=1}^K\bar{g}_k\Hm_k\Hm_k\htp\RB\xrightarrow[N\to\infty]{\text{a.s.}}0
\end{align*}
since $\lVert\sum_{k=1}^K(\bar{e}_k-\bar{g}_k)\Hm_k\Hm_k\htp\rVert\xrightarrow[]{\text{a.s.}}0.$ Applying Corollary~\ref{cor:logdet} to the second therm yields
\begin{align}\label{eq:conv3}
\frac1N\log\det\LB\Id_N+\frac{1}{\rho}\sum_{k=1}^K\bar{g}_k\Hm_k\Hm_k\htp\RB - \bar{\Vc}_N(\rho)\xrightarrow[N\to\infty]{\text{a.s.}}0.
\end{align}
Consider now $\bar{\Ic}_N(\rho)$ and $\bar{I}_N(\rho)$ as defined in Theorems~\ref{th:logdetdet} and \ref{th:mutinf}. It follows from \eqref{eq:conv1}, \eqref{eq:conv2} and \eqref{eq:conv3}, that
\begin{align*}
 \bar{\Ic}_N(\rho) - \bar{I}_N(\rho) \xrightarrow[N\to\infty]{\text{a.s.}}0.
\end{align*}
This implies also that
\begin{align}\label{eq:conv4}
I_N(\rho) - \bar{I}_N(\rho)\xrightarrow[N\to\infty]{\text{a.s.}}0.
\end{align}
To prove mean convergence $(ii)$, consider $\Omega$, the probability space that engenders the sequences $\{\Wm_1,\dots,$ $\Wm_K,\Hm_1,\dots,\Hm_K\}$. Then, on a subspace of $\Omega$ of measure $1$, we have by \eqref{eq:conv4}:  $I_N(\rho) - \bar{I}_N(\rho) \to 0$ as $N\to\infty$. The results follows directly by integrating this expression over $\Omega$, using the dominated convergence theorem \cite[Theorem 16.4]{billingsley}.
\end{IEEEproof}

\vspace{10pt}\begin{IEEEproof}[Proof of Proposition \ref{prop:pow}]
By the chain rule of differentiation, we have
\begin{align*}
 \frac{d\bar{I}_N(\rho)}{dp_{kj}}&= \frac{\partial\bar{I}_N(\rho)}{\partial p_{kj}} + \sum_{i=1}^K\LSB \frac{\partial\bar{I}_N(\rho)}{\partial g_{i}}\frac{\partial g_{i}}{\partial p_{kj}} +\frac{\partial\bar{I}_N(\rho)}{\partial \bar{g}_{i}}\frac{\bar{g}_{i}}{\partial p_{kj}}\RSB.
\end{align*}
 Consider now the partial derivative:
\begin{align*}
 \frac{\partial\bar{I}_N(\rho)}{\partial g_{i}} &=\frac{\partial \bar{V}_N(\rho)}{\partial g_{i}} + \frac1N\trace\Pm_i\LB g_i\Pm_i + \LSB\bar{c}_i-g_i\bar{g}_i\RSB\Id_{n_i}\RB^{-1}-\bar{g}_i\frac1N\sum_{j=1}^{n_i}\frac{1}{\bar{c}_i -g_i\bar{g}_i + g_i p_{ij}}-\bar{g}_i \frac{(1-c_i)\bar{c}_i}{\bar{c}_i-g_i\bar{g}_i}\\
&= \frac{\partial \bar{V}_N(\rho)}{\partial g_{i}} + \bar{g}_i \LB1-\frac{(1-c_i)\bar{c}_i}{\bar{c}_i-g_i\bar{g}_i} -\frac1N\sum_{j=1}^{n_i}\frac{1}{\bar{c}_i -g_i\bar{g}_i + g_i p_{ij}}\RB\\
& = \frac{\partial \bar{V}_N(\rho)}{\partial g_{i}}
\end{align*}
where the last equality follows from
\begin{align*}
 0 &= \bar{c}_i - \frac1N\sum_{j=1}^{n_i}\frac{\bar{c}_i-g_i\bar{g}_i + g_i p_{ij}}{\bar{c}_i-g_i\bar{g}_i + g_i p_{ij}} - \frac1N\sum_{j=1}^{N_i-n_i}\frac{\bar{c}_i-g_i\bar{g}_i}{\bar{c}_i-g_i\bar{g}_i}\\
& = \bar{c}_i - (\bar{c}_i-g_i\bar{g}_i) \frac1N\sum_{j=1}^{n_i}\frac{1}{\bar{c}_i-g_i\bar{g}_i + g_i p_{ij}} +  g_i\frac1N\sum_{j=1}^{n_i}\frac{p_{ij}}{\bar{c}_i-g_i\bar{g}_i + g_i p_{ij}} -(\bar{c}_i-g_i\bar{g}_i) \frac{\frac{N_i}{N}-\frac{n_i}{N}}{\bar{c}_i-g_i\bar{g}_i} \\
& = (\bar{c}_i-g_i\bar{g}_i)\LB1 - \frac{(1-c_i)\bar{c}_i}{\bar{c}_i-g_i\bar{g}_i}- \frac1N\sum_{j=1}^{n_i}\frac{1}{\bar{c}_i-g_i\bar{g}_i + g_i p_{ij}}\RB
\end{align*}
and $\bar{c}_i\ge\bar{c}_i c_i> g_i\bar{g}_i$ by definition.

Similarly, we have
\begin{align*}
  \frac{\partial\bar{I}_N(\rho)}{\partial \bar{g}_{i}} &=\frac{\partial \bar{V}_N(\rho)}{\partial \bar{g}_{i}} - g_i\LB\frac{(1-c_i)\bar{c}_i}{\bar{c}_i-g_i\bar{g}_i}+ \frac1N\sum_{j=1}^{n_i}\frac{1}{\bar{c}_i-g_i\bar{g}_i + g_i p_{ij}}\RB \\
&=\frac{\partial \bar{V}_N(\rho)}{\partial \bar{g}_{i}} - g_i.
\end{align*}

It remains now to calculate the partial derivatives $\frac{\partial \bar{V}_N(\rho)}{\partial g_{i}}$ and $\frac{\partial \bar{V}_N(\rho)}{\partial \bar{g}_{i}}$.  To this end, notice that
\begin{align*}
 1=\frac1N\trace\Tm\Tm^{-1} &= \rho\frac1N\trace\Tm + \sum_{k=1}^K\bar{g}_k\frac1N\sum_{j=1}^{N_k} \frac{\frac1N\trace{\Rm_{k,j}\Tm}}{1+\bar{g}_k\frac1N\trace\Rm_{k,j}\Tm}\\
&=  \rho\frac1N\trace\Tm + \sum_{k=1}^K\bar{g}_kg_k.
\end{align*}
Replacing $\sum_{k=1}^K\bar{g}_kg_k$ in \eqref{eq:mutinfeq} by $\frac1N\trace\Tm\Tm^{-1} - \rho\frac1N\trace\Tm$ yields
\begin{align*}
\bar{V}_N(\rho)= -\frac1N\log\det\LB\rho\Tm\RB-\frac1N\trace\Tm\Tm^{-1} +\rho\frac1N\trace\Tm  + \frac1N\sum_{k=1}^K\sum_{j=1}^{N_k}\log\LB1+\bar{g}_k\frac{1}{N}\trace\Rm_{k,j}\Tm\RB.
\end{align*}
Taking the derivative with respect to $\bar{g}_i$ and denoting $\Tm'= \frac{\partial \Tm}{\partial \bar{g}_{i}}$ leads to
\begin{align*}
 \frac{\partial \bar{V}_N(\rho)}{\partial \bar{g}_{i}} & = -\frac1N\trace\Tm^{-1}\Tm' + \rho\frac1N\trace\Tm' + \frac1N\sum_{k=1}^K\sum_{j=1}^{N_k}\frac{\bar{g}_k\frac{1}{N}\trace\Rm_{k,j}\Tm'}{1+\bar{g}_k\frac{1}{N}\trace\Rm_{k,j}\Tm} + \frac1N\sum_{j=1}^{N_i}\frac{\frac{1}{N}\trace\Rm_{i,j}\Tm}{1+\bar{g}_i\frac{1}{N}\trace\Rm_{i,j}\Tm}\\
& = -\rho\frac1N\trace\Tm' - \frac1N\sum_{k=1}^K\sum_{j=1}^{N_k}\frac{\bar{g}_k\frac{1}{N}\trace\Rm_{k,j}\Tm'}{1+\bar{g}_k\frac{1}{N}\trace\Rm_{k,j}\Tm} + \rho\frac1N\trace\Tm' + \frac1N\sum_{k=1}^K\sum_{j=1}^{N_k}\frac{\bar{g}_k\frac{1}{N}\trace\Rm_{k,j}\Tm'}{1+\bar{g}_k\frac{1}{N}\trace\Rm_{k,j}\Tm} + g_i\\
& = g_i.
\end{align*}
This implies that $\frac{\partial\bar{I}_N(\rho)}{\partial \bar{g}_{i}}=0$. 
We similarly have 
\begin{align*}
 \frac{\partial \bar{V}_N(\rho)}{\partial g_{i}} = 0
\end{align*}
and hence $\frac{\partial\bar{I}_N(\rho)}{\partial g_{i}}=0$. 
Putting the last results together yields 
\begin{align}\label{eq:dermutint}
  \frac{d\bar{I}_N(\rho)}{dp_{kj}}&= \frac{\partial\bar{I}_N(\rho)}{\partial p_{kj}}=\frac{g_k}{N\LB\bar{c}_k-g_k\bar{g}_k + g_kp_{kj}\RB}.
\end{align}
We can calculate the second derivative in a similar manner:
\begin{align*}
  \frac{d^2\bar{I}_N(\rho)}{dp_{kj}^2}&= \frac{\partial^2\bar{I}_N(\rho)}{\partial p_{kj}^2} =-\frac{g_k^2}{N\LB\bar{c}_k-g_k\bar{g}_k + g_kp_{kj}\RB^2}\le0
\end{align*}
since $g_k\ge0$. Thus, $\bar{I}_N(\rho)$ is a concave function in $p_{kj}$ for all $k,j$. It is straightforward to verify that also $I_N(\rho)$ is concave in all $p_{kj}$.

Consider now the Lagrangian functions related to the power constraints $(I)$ and $(II)$:
\begin{align}
 \Lc(\lambda,\lambda_1,\dots,\lambda_K,p_{11},\dots,p_{KN_K})= \begin{cases} \bar{I}_N(\rho) - \sum_{k=1}^K\lambda_k\LB\frac1{n_k}\sum_{j=1}^{n_k} p_{kj}-P_k\RB & (I)\\
                                                  \bar{I}_N(\rho) - \lambda\LB\sum_{k=1}^K\frac1{n_k}\sum_{j=1}^{n_k} p_{kj}-P\RB & (II)
                                                 \end{cases}
\end{align}
We have from \eqref{eq:dermutint}
\begin{align}\label{eq:lagrangeder}
 \frac{\partial \Lc}{\partial p_{kj}} = \begin{cases}\frac{g_k}{N\LB\bar{c}_k-g_k\bar{g}_k + g_kp_{kj}\RB} - \frac{\lambda_k}{n_k} & (I)\\
                           \frac{g_k}{N\LB\bar{c}_k-g_k\bar{g}_k + g_kp_{kj}\RB} - \frac{\lambda}{n_k} & (II)
                          \end{cases}.
\end{align}
Solving for the Karush-Kuhn-Tucker conditions \cite{boyd_cvx} for both cases yields the desired result.

Take now the optimal solutions $\bar{\Pm}^*\defines(\bar{\Pm}_1^*,\dots,\bar{\Pm}_K^*)$ and $\Pm^*\defines(\Pm_1^*,\dots,\Pm_K^*)$ and consider the following difference:
\begin{align*}
 I_N(\Pm^*) - I_N(\bar{\Pm}^*) & =  \LSB I_N(\Pm^*) -\bar{I}_N(\Pm^*)\RSB + \LSB\bar{I}_N(\Pm^*)  - \bar{I}_N(\bar{\Pm}^*)\RSB + \LSB\bar{I}_N(\bar{\Pm}^*)-I_N(\bar{\Pm}^*)\RSB
\end{align*}
where we used $I_N(\Pm^*)$ and $\bar{I}_N(\bar{\Pm}^*)$ to denote $I_N(\rho)$ and $ \bar{I}_N(\rho)$ evaluated for the matrices $(\bar{\Pm}_1^*,\dots,\bar{\Pm}_K^*)$ and $(\Pm_1^*,\dots,\Pm_K^*)$ , respectively.
Assuming that $\max_K\lim\sup_N\lVert\Pm^*_k\rVert\le \infty$, we have from Theorem~\ref{th:mutinf}
\begin{align*}
 I_N(\Pm^*) -\bar{I}_N(\Pm^*) &\xrightarrow[N\to\infty]{\text{a.s.}} 0\\
\bar{I}_N(\bar{\Pm}^*)-I_N(\bar{\Pm}^*)&\xrightarrow[N\to\infty]{\text{a.s.}} 0.
\end{align*}
Since $I_N(\Pm^*) - I_N(\bar{\Pm}^*)\ge0$ and $\bar{I}_N(\Pm^*)-\bar{I}_N(\bar{\Pm}^*)\le 0$, we can conclude that
\begin{align*}
  I_N(\Pm^*) - I_N(\bar{\Pm}^*)\xrightarrow[N\to\infty]{\text{a.s.}} 0.
\end{align*}
It remains now to show that the matrices $\Pm^*_k$ satisfy indeed $\max_K\lim\sup_N\lVert\Pm^*_k\rVert\le \infty$. Consider therefore the following expression:
\begin{align*}
 \mathbb{E}I_N & = \mathbb{E}\frac1N\log\det\LB\Id_N + \frac1\rho{\Bm_N}_{[kj]}+\frac{p_{kj}}{\rho}\Hm_k\wv_{kj}\wv_{kj}\htp\Hm_k\htp\RB
\end{align*}
which is clearly strictly concave in $p_{kj}$ for all $k,j$. The corresponding derivative with respect to $p_{kj}$ reads
\begin{align*}
\frac{\partial\mathbb{E}I_N}{\partial p_{kj}} & = \mathbb{E}\frac1N\trace\LB\Id_N+\frac1\rho\Bm_N\RB^{-1}\frac1\rho\Hm_k\wv_{kj}\wv_{kj}\htp\Hm_k\htp\\
& = \mathbb{E}\frac{1}{\rho N} \wv_{kj}\htp\Hm_k\htp\LB\Id_N+\frac1\rho\Bm_N\RB^{-1}\Hm_k\wv_{kj}.
\end{align*}
Similar to \eqref{eq:lagrangeder}, the derivative of the Lagrangian to the optimization problem \eqref{eq:optIn} is given as
\begin{align*}
 \frac{\partial \Lc}{\partial p_{kj}} = \begin{cases}\mathbb{E}\frac{1}{\rho N} \wv_{kj}\htp\Hm_k\htp\LB\Id_N+\frac1\rho\Bm_N\RB^{-1}\Hm_k\wv_{kj} - \frac{\lambda_k}{n_k} & (I)\\
                            \mathbb{E}\frac{1}{\rho N} \wv_{kj}\htp\Hm_k\htp\LB\Id_N+\frac1\rho\Bm_N\RB^{-1}\Hm_k\wv_{kj}- \frac{\lambda}{n_k} & (II)
                          \end{cases}.
\end{align*}
Consider now constraint $(I)$. At the optimal point, we need to have $\frac{\partial \Lc}{\partial p_{kj}}=0$, and therefore
\begin{align*}
 \mathbb{E}\frac{1}{\rho N} \wv_{kj}\htp\Hm_k\htp\LB\Id_N+\frac1\rho\Bm_N\RB^{-1}\Hm_k\wv_{kj} = \frac{\lambda_k}{n_k}.
\end{align*}
Since the right-hand side is independent of $j$, it follows that $\Pm^*_k = p_k\Id_{n_k}$ where $p_k$ is a parameter to be optimized. Since $\frac1{n_k}\trace\Pm_k^*=p_k\le P_k$, we have $\max_K\lim\sup_N\lVert\Pm^*_k\rVert=p_k\le P_k <\infty$. The same arguments hold for the sum-power constraint $(II)$.
\end{IEEEproof}

\vspace{20pt}\begin{IEEEproof}[Proof of Theorem \ref{th:sinr}]
 The proof follows directly from \eqref{eq:conv1} and \eqref{eq:conv2} applied to \cite[Theorem 4]{couillet10b}.
\end{IEEEproof}

\vspace{20pt}\begin{IEEEproof}[Proof of Corollary \ref{cor:MMSE}]
 The proof of $(i)$ follows directly from the continuous mapping theorem \cite[Theorem 2.3]{vdv}. Denote $\Omega$ the probability space engendering the sequences $\{\Wm_1,\dots,\Wm_K,\Hm_1,\dots,\Hm_K\}$. Then, on a sub-space of $\Omega$ of measure $1$, we have by Theorem~\ref{th:mutinf}: $R_N(\rho) - \bar{R}_N(\rho) \to 0$ as $N\to\infty$. Integrating this expression over $\Omega$ using dominated convergence arguments proves $(ii)$.
\end{IEEEproof}

\section{Related results}\label{sec:appendixB}
\begin{theorem}[{\cite[Theorem 1]{wagner2011}}]\label{th:detequcorr}
 Let $\Bm_N=\Xm\Xm\htp$, where $\Xm\in\CC^{N\times n}$ is random. The $j$th column $\xv_{j}$ of $\Xm$ is given as $\xv_{j} = \Rm_j^{\frac12} \uv_{j}$, where the entries of $\uv_{j}\in\CC^{N}$ are i.i.d.\@ with zero mean, variance $1/N$ and finite moment of order $4+\epsilon$, for some common $\epsilon>0$, and $\Rm_{j}\in\CC^{N\times N}$ are Hermitian nonnegative definite matrices. Let $\Dm_N\in\CC^{N\times N}$ be a deterministic Hermitian matrix. Assume that both $\Rm_{j}$ and $\Dm_N$ have uniformly bounded spectral norms (with respect to $N$). Then, as $n,N\to\infty$ such that $0< \lim\inf N/n \le \lim\sup N/n < \infty$, the following holds for any $z\in\CC\setminus\RR_+$:
\begin{align*}
 \frac1N\trace\Dm_N\LB\Bm_N-z\Id_N\RB^{-1} - \frac1N\trace\Dm_N\Tm_N(z) \xrightarrow[N,n\to\infty]{\text{a.s.}}0
\end{align*}
where $\Tm_N(z)\in\CC^{N\times N}$ is defined as
\begin{align*}
 \Tm_N(z) = \LB\frac1N\sum_{j=1}^n\frac{\Rm_{j}}{1+\delta_{j}(z)}-z\Id_N\RB^{-1}
\end{align*}
and where $\delta_{1}(z),\dots,\delta_{n}(z)$ are given as the unique solution to the following set of implicit equations:
\begin{align}\label{eq:fixedpoint}
 \delta_{j}(z)=\frac1N\trace\Rm_{j}\LB\frac1N\sum_{j=1}^n\frac{\Rm_{j}}{1+\delta_{j}(z)}-z\Id_N\RB^{-1},\qquad j=1,\dots,n
\end{align}
such that $(\delta_{1}(z),\dots,\delta_{n}(z))\in\Sc^n$. For $z<0$, $\delta_{1}(z),\dots,\delta_{N,n}(z)$ are the unique nonnegative solutions to \eqref{eq:fixedpoint} and can be obtained by a standard fixed-point algorithm with initial values $\delta_{j}^{(0)}(z)=-1/z$ for $j=1,\dots,n$.
 Moreover, let $F_N$ be the empirical spectral distribution (e.s.d.) of $\Bm_N$ and denote by $\bar{F}_N$ the distribution function with Stieltjes transform $\frac1N\trace\Tm_N(z)$. Then, almost surely,
\begin{align*}
 F_N - \bar{F}_N \Rightarrow 0.
\end{align*}
\end{theorem}

\vspace{10pt}\begin{theorem}[{\cite{wagnerphd}}]\label{th:logdetcor}
 Under the assumptions of Theorem~\ref{th:detequcorr}, let $\rho>0$ and define the quantity $\Vc_N(\rho)=\frac1N\log\det\LB\Id_N+\frac{1}{\rho}\Bm_N\RB$. Then,
\begin{align*}
 \mathbb{E}\Vc_N(\rho) - \bar{\Vc}_N(\rho) \xrightarrow[N,n\to\infty]{} 0
\end{align*}
where
\begin{align*}
 \bar{\Vc}_N(\rho) &= \frac1N\log\det\LB\Id_N  +  \frac{1}{\rho}\frac1N\sum_{j=1}^n\frac{\Rm_{j}}{1+\delta_{j}}\RB + \frac1N\sum_{j=1}^n\log\LB1+\delta_{j}\RB - \frac1N\sum_{j=1}^n \frac{\delta_{j}}{1+\delta_{j}}
\end{align*}
and where $\delta_{j}=\delta_{j}(-\rho)$ for $j=1,\dots,n$ are given by Theorem~\ref{th:detequcorr}.
\end{theorem}

\vspace{10pt}\begin{cor}\label{cor:logdet}
 Under the assumptions of Theorem~\ref{th:logdetcor}, assume additionally that the matrices $\Rm_{j}$, $j=1,\dots,n$, are drawn from a finite set of Hermitian nonnegative-definite matrices. Then, 
\begin{align}
\Vc_N(\rho) - \bar{\Vc}_N(\rho) \xrightarrow[N,n\to\infty]{\text{a.s.}} 0
\end{align}
where $\Vc_N(\rho)$ and $\bar{\Vc}_N(\rho)$ are defined as in Theorem~\ref{th:logdetcor}.
\end{cor}
\begin{IEEEproof}
 It was shown in \cite[Proof of Theorem 3]{hoydis2011} that $\Bm_N$ has almost surely uniformly bounded spectral norm as $N,n\to\infty$ if the matrices $\Rm_{j}$ are drawn from a finite set of matrices. Thus, $F_N$ and $\bar{F}_N$ as defined in Theorem~\ref{th:detequcorr} have (almost surely) bounded support. Consider now a probability space $A\subset\Omega$, $\Omega$ generating the matrices $\Bm_N$, for which $\Bm_N$ has bounded spectral norm, and another probability space $B\subset\Omega$ for which $F_N - \bar{F}_N \Rightarrow 0$. Since $P(A)=P(B)=P(A\cap B)=1$, it follows from \cite[Theorem 25.8 (ii)]{billingsley}, that
\begin{align}
 \int\log(1+x\lambda)dF_N(\lambda) - \int\log(1+x\lambda)d\overline{F}_N(\lambda)\xrightarrow[N,n\to\infty]{\text{a.s.}} 0
\end{align}
which is equivalent to stating that $\Vc_N(x) - \bar{\Vc}_N(x) \xrightarrow[N,n\to\infty]{\text{a.s.}} 0$.
\end{IEEEproof}

\section{Useful Lemmas}
\begin{lemma}[Resolvent identity]\label{lem:resolvent} Let $\Am$ and $\Bm$ be two invertible matrices. Then,
\begin{align}
 \Am^{-1}-\Bm^{-1} = \Am^{-1}\LB\Bm-\Am\RB\Bm^{-1}.
\end{align}
\end{lemma}

\begin{lemma}[Matrix inversion lemma {\cite[Eq. (2.2)]{silverstein95}}]\label{lem:inversion}
 Let $\Am\in\CC^{N\times N}$ be Hermitian invertible. Then, for any vector $\xv\in\CC^N$ and any scalar $\tau\in\CC$ such that $\Am+\tau \xv\xv\htp$ is invertible,
\begin{align}
 \xv\htp \LB\Am+\tau \xv\xv\htp\RB^{-1} = \frac{\xv\htp\Am^{-1}}{1+\tau \xv\htp \Am^{-1}\xv}. 
\end{align}
\end{lemma}

\begin{lemma}[Rank-$1$ perturbation lemma \cite{silverstein95}]\label{lem:rank1perturbation}
Let $z<0$, $\Am\in\CC^{N\times N}$, $\Bm\in\CC^{N\times N}$ with $\Bm$ Hermitian nonnegative definite, and $\vv\in \CC^N$. Then,
\begin{align}
    \left|\trace\LB(\Bm-z\Id_N)^{-1}-(\Bm+\vv\vv\htp-z\Id_N)^{-1}\RB\Am\right|\leq\frac{\Vert \Am \Vert}{|z|}\ . 
\end{align}
\end{lemma} 

\begin{lemma}[Trace lemma {\cite[Lemma 2.7]{silverstein98}}]\label{lem:trace}
 Let $\Am_1,\Am_2,\dots$, with $\Am_N\in\CC^{N\times N}$, be a sequence of matrices with uniformly bounded spectral norm and let $\xv_N=\in\CC^N$ be random vectors of i.i.d.\@ entries with zero mean, variance $1/N$ and eighth order moment of order $\Oc(1/N^4)$, independent of $\Am_N$. Then,
\begin{align}
 \xv_N\htp\Am_N\xv_N-\frac1N\trace\Am_N\xrightarrow[N\to\infty]{\text{a.s.}}0.
\end{align}
\end{lemma}

\begin{lemma}\cite[Lemma 1]{peacock08}\label{lem:convergence_ratios}
Denote $a_N$, $\overline{a}_N$, $b_N$ and $\overline{b}_N$ four infinite sequences of complex random variables indexed by $N$ and assume $a_N\asymp \overline{a}_N$ and $b_N\asymp \overline{b}_N$. If $|a_N|$, $|\overline{b}_N|$ and/or $|\overline{a}_N|$,$|b_N|$ are uniformly bounded above over $N$ (almost surely), then $a_Nb_N\asymp \overline{a}_N \overline{b}_N$. Similarly, if $|a_N|$, $|\overline{b}_N|^{-1}$ and/or $|\overline{a}_N|$,$|b_N|^{-1}$ are uniformly bounded above over $N$ (almost surely), then $a_N/b_N\asymp \overline{a}_N/ \overline{b}_N$.
\end{lemma}

\begin{lemma}[Trace inequality]\label{lem:traceinequ}
 Let $\Am,\Bm,\Rm\in\CC^{N\times N}$, where $\Am$ and $\Bm$ are nonnegative-definite, satisfying $\Bm\succ \Am$, and $\Rm$ is nonnegative-definite. Then
\begin{align}
 \trace\Rm\LB\Am^{-1} - \Bm^{-1}\RB > 0.
\end{align}
\end{lemma}
\begin{IEEEproof}
Note that $\Bm\succ\Am$ implies by \cite[Corollary 7.7.4]{Horn} $\Bm^{-1}\prec\Am^{-1}$. Thus, for any vector $\xv\in\CC^N$,
\begin{align}
 \xv\htp\LB\Am^{-1}-\Bm^{-1}\RB\xv> 0 .
\end{align}
Consider now the eigenvalue decomposition of the matrix $\Rm=\Um\Lambdam\Um\htp$, where $\Um=\LSB\uv_1,\dots,\uv_{N}\RSB$ and $\Lambdam=\diag(\lambda_1,\dots,\lambda_{N})$. Since $\lambda_i\ge 0\ \forall i$, we have 
\begin{align}
 \trace\Rm\LB\Am^{-1} - \Bm^{-1}\RB &= \sum_{i=1}^N \lambda_i \uv_i\htp\LB\Am^{-1} - \Bm^{-1}\RB\uv_i > 0.
\end{align}
\end{IEEEproof}

\bibliographystyle{IEEEtran}
\bibliography{IEEEabrv,bibliography}
\end{document}